\documentclass[9.5pt,letter,twoside]{rho-class/rho}
\usepackage[english]{babel}
\pdfoutput=1

\usepackage{tikz}
\usepackage{amsmath} 
\usepackage{graphicx}
\usepackage{float}
\usepackage{booktabs}
\usepackage{enumitem}
\usepackage{xcolor}   
\usepackage{natbib}
\usepackage{pgfplots}
\pgfplotsset{compat=1.17}
\usepackage{tabularx}
\usepackage{hyperref}

\definecolor{possibilityColor}{RGB}{0, 102, 204}  
\definecolor{necessityColor}{RGB}{204, 0, 0}      

\newcommand*{\clyfar}{\textsc{Clyfar}}

\newenvironment{axiomlist}
    {\begin{list}{}{\setlength{\leftmargin}{0pt}\setlength{\itemsep}{6pt}\setlength{\parsep}{3pt}}}
    {\end{list}}

\newcommand{\nec}{$\text{N}$}
\newcommand{\poss}{$\Pi$}

\setbool{rho-abstract}{true} 
\setbool{corres-info}{true} 

\journalname{A Preprint}
\title{Communicating Risk with Possibility, Not Probability}

\author[1,2]{John R.\ Lawson}

\affil[1]{Department of Mathematics and Statistics, Utah State University, Vernal, Utah, United States of America}
\affil[2]{Bingham Research Center, Utah State University, Logan, Utah, United States of America}

\dates{Draft date: \today}

\leadauthor{John R.\ Lawson}
\footinfo{Communicating Risk with Possibility, Not Probability}
\smalltitle{\LaTeX\ Preprint}
\institution{Bingham Research Center, Utah State University}
\theday{\today\newline{}Submitted to XXXX}

\corres{Correspondence to John R.\ Lawson}
\email{john.lawson@usu.edu}

\license{Rho LaTeX Class \ccLogo\ This document is licensed under Creative Commons CC BY 4.0. Template further modified by the author.}

\begin{abstract}
Communicating forecast uncertainty effectively is a persistent challenge in predictive endeavours such as weather forecasting. This paper explores the application of possibility theory as a complementary approach to traditional probability in risk communication. Unlike probability, possibility theory allows for the representation of uncertain events as ranges of potential, distinguished by degrees of plausibility (\textit{possibility}) and certainty (\textit{necessity}). Using a simplified fuzzy-logic inference system, we generate possibility forecasts of ozone-concentration forecasts from meteorological inputs. Observations are pre-processed as degrees of an adjective; e.g., 1\,$ms^{-1}$ wind speed may belong to the adjective ``calm" at degree 0.8 (perhaps represented with an adverb as ``substantially"). Aggregation of all rule activations yields a possibility distribution over the output range. As possibility is an upper bound of probability, possibilities provides most value for risk-averse stakeholders sensitive to the event of interest, especially near the event's predictability horizon when there is most uncertainty to remove. By setting the forecast challenge as the possibility of an event, we effectively extend the predictability horizon: trading some specificity for detection of fainter signals of the event's future occurrence. Possibility theory appreciates uncertainty’s dual nature: inherent randomness (\textit{aleatoric}) and knowledge deficiency (\textit{epistemic}). While the unfamiliar nature of the theory requires more accessible language before operational use and public communication, \textit{possibility} offers substantial practical benefits for assessment of uncertainty at the edge of predictability limits for vulnerable users.
\end{abstract}

\keywords{possibility, probability, uncertainty, risk communication, fuzzy logic, air quality, ozone}

\begin{document}
    \maketitle
    \thispagestyle{firststyle}
    \tableofcontents



\section*{Significance Statement}
\begingroup
\hyphenpenalty=10000
\exhyphenpenalty=10000
\begin{itemize}
    \item Decision-makers benefit from understandable expression of forecast uncertainty, typically given as probabilities. 
    \item Possibility theory gives an alternative measure of confidence in event $A$ occurring, comprising the plausibility of $A$, \textit{possibility}, and its certainty or \textit{necessity}.
    \item It is easier to predict $A$ is \textit{possible} rather than \textit{probable}, extending the predictability time horizon and providing valuable information to risk-averse stakeholders for low-predictability events.
    \item We demonstrate the method of generating possibilities of future events from a simplified fuzzy-logic inference system.
    \item The paradigm improves rare-event risk communication leveraging the idea of confidence in confidence: when inevitable, $A$ is fully \textit{possible} and entirely \textit{necessary}.
\end{itemize} 
\endgroup


\section{Introduction}\label{sec:intro}
\begin{figure}[H]
    \centering
    \includegraphics[width=\columnwidth]{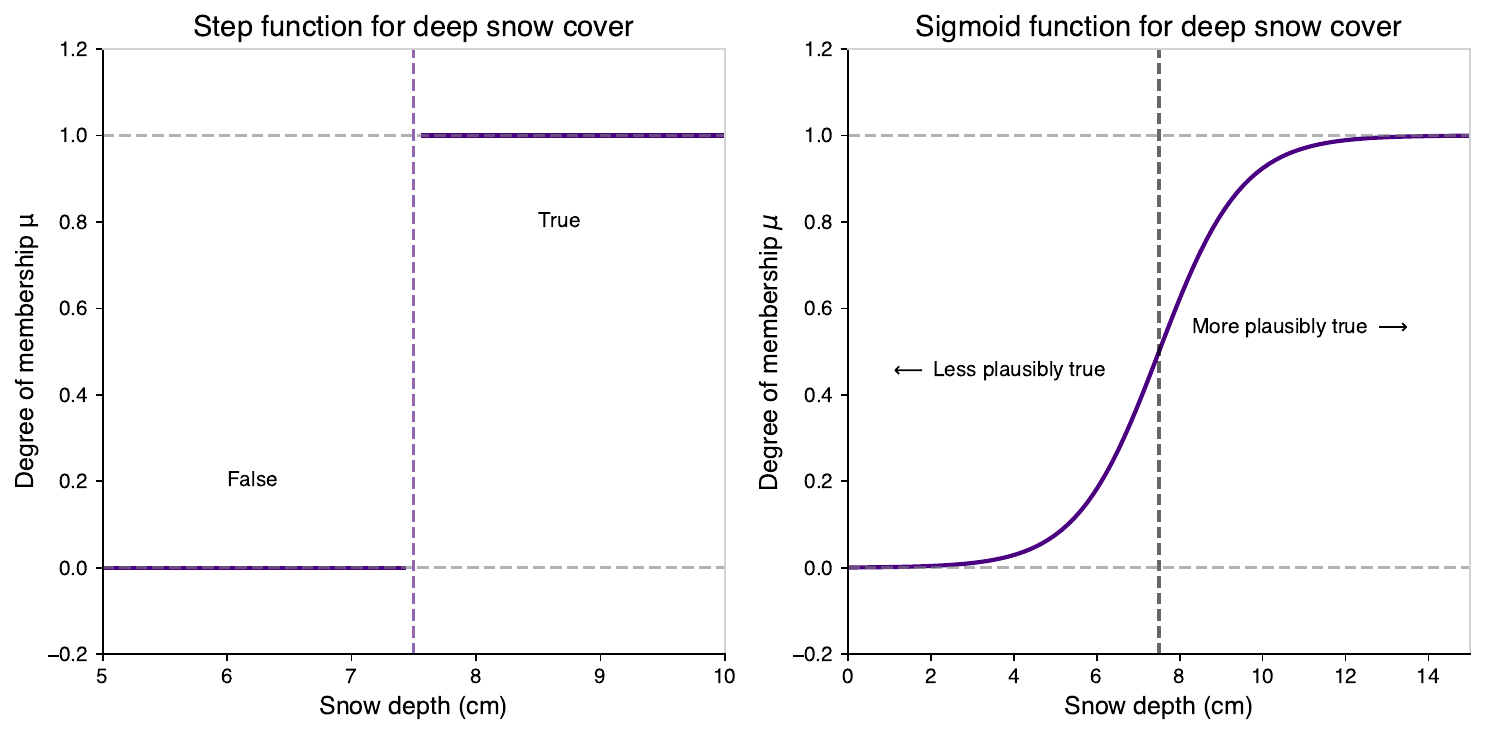}
    \caption{Bivalent versus fuzzy logic for the example of snow depth: ``how deep is \textit{deep}?".}
    \label{fig:twologics}
\end{figure}

\rhostart{U}ncertainty of predictions is difficult but essential to communicate \citep[e.g.,][]{Murphy1977-sm,Jardine1997-tn,Gigerenzer2005-zs,Joslyn2009-rg,Ahmad2015-aw,Juanchich2019-ne,Qin2023-oy}. In the U.S., the National Weather Service (NWS) is embracing probabilistic thinking to incorporate and guide best practices in modelling \citep{Hirschberg2011-sl,Tripp2022-jf,Roberts2019-ox,Flora2024-yt} and communication \citep{Rothfusz2018-yk,Klockow-McClain2019-zs,Gerst2020-rg,Trujillo-Falcon2022-du}. However, the challenge is compounded when probabilistic forecasts are uncertain themselves: a second-order or ``meta-probability" \citep{Fraedrich1987-im,Palmer2000-kk,Van-Schaeybroeck2016-sl,Coleman2024-hf}. Consider risk-averse users whose sensitivity to a damaging event grows quickly as the probability of an event increases gradually. These cautious users would benefit from knowing their issued event risk has a wide bound of uncertainty and is therefore potentially more likely than forecast. Conceptually, just as calibration and sharpness must be optimised to maximise information gained by a forecast \citep{Hersbach2000-yb,Lawson2024-bu}, if there is uncertainty in uncertainty, there is an optimal ``sharp" likelihood within the \textit{fuzzy} probability spanning upper and lower bounds. Fuzzy logic is fundamental to possibility theory: an example of \textit{truthiness} (Fig.~\ref{fig:twologics}): the degree of truth that a snow depth of 9.3\,cm is ``deep", labelled "True" for two-valued (bivalent) and fuzzy logics. 

\subsection*{Alternative theories of uncertainty}
Bounded (uncertain) probabilities have emerged independently in the literature under multiple names. Alternative theories of uncertainty have been proposed to address the limitations of traditional probability theory in representing uncertainty. It is difficult to determine the probability of a rare event through observed frequency with insufficiently few archived observations. For instance, ``50\%" could be issued with confidence if half of computer simulations predict event $A$; however, an unprecedented event $A^!$ may draw a na\"ive 50\% prior in the absence of further information. Two related paradigms address this by splitting probability into dual concepts that bound uncertainty in $A$. Dempster-Shafer (D--S) theory assigns \textit{belief} and \textit{plausibility} measures (summarised in \citealt{Yager2005-at}), analogous to upper and lower bounds of probabilities, or \textit{possibility} and \textit{necessity} in possibility theory \citep{Dubois1992-gd}. As the root of possibility theory, fuzzy logic offers a nuanced approach by handling partial memberships. It is the ability to encode expert knowledge as fuzzy logic \citep{Zadeh1965-wi} that fuels the author's preference for a \textit{possibility-theory} paradigm over alternatives (e.g., D--S theory), further to persuasive arguments in \citet{Dubois2010-fe}.

\subsection*{Sampling time--space and uncertainty dimensions}
Predicting atmospheric states is typically performed with numerical weather prediction (NWP) models that integrate equations of motion on grids for a global or limited domain. Given finite computer power, there is an optimisation problem when sampling time, space, and uncertainty dimensions: the trade-off between NWP grid resolution and ensemble membership. When halving horizontal grid spacing (e.g., $\Delta x = 2\,\text{km}$ to 1\,km), compute requirements increase by \textit{O}(10)---neglecting system bottlenecks, and whilst keeping ensemble membership identical (see Appendix~\ref{appx:mfs}). For a meteorological example, fine resolution ($<3\,\text{km}$) is required for adequate simulation of thunderstorms \citep{Schwartz2019-js} and complex mountainous atmospheric structure \citep{Doyle2000-rj,Neemann2015-ws}. This problem is generalised, however, as aiming to remove prior uncertainty that grows as the question becomes more complex (i.e., there are more degrees of freedom in this subset $\Omega_0$). \textbf{It is an easier question to ask if $A^!$ is possible than probable}. There is high value if warning can be given (information gained or surprise removed) as early as possible to risk-averse users. Estimating the possibility of a high-impact event $A^!$, while not as immediately actionable as probability, effectively extends the predictability horizon \citep{Lorenz1963-zy} in an information-gain sense when forecasts of $A^!$ reveal no further information, on average, over a baseline such as its base-rate. 

\subsection*{Objective and Scope}
This paper explores whether possibility theory can extend predictability horizons for risk-averse users in a way probability theory cannot, and whether these insights are both useful and accessible for these highly sensitive forecast users. Additionally, we discuss the implications of normalising possibility distributions and impact on the representation of uncertainty. Our focus is in an atmospheric-science setting but applies to all predictive fields.

For illustrative examples, we employ simplified elements from a fuzzy inference system (FIS) in \citet{Lawson2024-jb} at the core of the ozone prediction model \clyfar. From this quasi-operational environment, we provide examples of the mathematical concepts of possibility and fuzzy logic. More information about the Bingham Research Center's \textit{Ozone Alert} program can be found at \url{https://www.usu.edu/binghamresearch/} (accessed 1 October 2024), while operational forecasts are available alongside live observations of air quality at \url{ubair.usu.edu} (accessed 1 October 2024).

\section{Background}\label{sec:background}
A drawback of probabilities is their requirement for a long record of rare events to determine their frequency, leading to ``black swans" \citep{Taleb2007-mt}: events (or swans \footnote[2]{Roman satirist Juvenal deployed the concepts of black swans as an \textit{adynaton}: something previously considered impossible or nonsensical. Black swans were long known to the native peoples of Australasia but mere myth to Dutch explorers until formal cataloguing by John Latham \citep{Latham1790-vj}.}) that are either unprecedented or have not been imagined. Hence it is sensible to acknowledge uncertainty from numerous sources, including inherent noise in the system, or our limited knowledge. In this light, we begin discussion of uncertainty by decomposing its concept into into two sorts: \textit{aleatoric} and \textit{epistolic}.

Aleatoric uncertainty is inherent, appearing as randomness or noise. This may stem from the chaotic nature of complex system \citep{Lorenz1963-uv,Williams1997-jm}, for instance. Conversely, epistemic uncertainty is a lack of understanding, whether it be system dynamics, initial conditions, etc. It is important to differentiate as aleatoric uncertainty cannot be reduced, only considered by, e.g., using risk-based forecasts such as the micromort (1-in-million chance of fatality per activity) popular in health sciences \citep{Blastland2014-wn,Ahmad2015-aw}. Observation and model improvements \textit{can} reduce epistolic error, however, whether it be meteorological \citep{Whitaker2012-en} or medical \citep{Picone2017-ql}.


\subsection{Fuzzy Logic}\label{sec:fuzzy_logic}
Probability theory is based on bivalent (two-valued) logic, where set membership is either True or False. In contrast, fuzzy logic \citep{Zadeh1965-wi} allows for degrees of membership, enabling more nuanced representations of uncertainty. The degree to which an element is part of a fuzzy set is determined by membership functions. An example of \textit{truthiness} in Figure~\ref{fig:twologics} contrasts bivalent logic (it \textbf{is}, or it \textbf{is not}) and fuzzy logic (there are degrees of truth) for snow depth being classed as \textit{deep} from its numerical measurement. For continuous quantities, such as determining if snow is deep, it is intuitive there is not a discontinuity (Fig.~\ref{fig:twologics}a) but a smooth transition around an ill-defined region or inflexion point (Fig.~\ref{fig:twologics}b). 


\subsection{Possibility Theory}\label{sec:poss_theory}
Possibility theory extends fuzzy logic by introducing dual measures: possibility \poss{} and necessity \nec.~\footnote[3]{Mirroring the convention for possibility, this is capital $\nu$, with the latter lower case used herein for a \textit{necessity distribution}.} We may consider synonyms for \poss{} such as plausibility or potential (but certainly not probability!), with terms like ``certainty" and ``inevitability" for \nec.~\footnote[4]{Another distinction is that possibility is generated by the support in data, i.e., the maximum of rule activations, as discussed in later Sections.} This framework allows for a more comprehensive representation of uncertainty compared with probability theory, sacrificing specificity (a confident probability value) for insight into both aleatoric and epistolic uncertainty.

One can define a membership function $\mu$ through human design \citep[p.59]{Dubois1988-nh} advised by experts and historical observation; forecasts herein are created via output from a fuzzy-logic inference system. To begin, we assert our forecast system's output membership function $\mu_o = \pi$: the possibility distribution. For all values in the variable range (alternatively worded as all events in the universe of discourse $\forall \omega \in \Omega$), there is a possibility value \poss{} that denotes plausibility of the event of interest $\omega = A$ occurring. Representing lexical range, we might convert values to/from adverbs and adjectives given approximate possibility ranges as in Table~\ref{tab:fuzzy_terms}.

\begin{table}[h!]
    \centering
    \caption{Fuzzy Logic Membership $\mu$ for Snow Depth labelled as \textit{deep}.}
    \label{tab:fuzzy_terms}
    \begin{tabular}{@{}cc@{}}
        \toprule
        \textbf{Membership $\mu$} & \textbf{Descriptor} \\ 
        \midrule
        $\mu = 0.0$ & Not at all deep \\
        $0 < \mu \leq 0.2$ & A little deep \\
        $0.2 < \mu \leq 0.4$ & Somewhat deep \\
        $0.4 < \mu \leq 0.6$ & Pretty deep \\
        $0.6 < \mu \leq 0.8$ & Substantially deep \\
        $\mu = 1.0$ & Absolutely deep \\
        \bottomrule
    \end{tabular}
    \vspace{0.5em}
\end{table}

This paradigm embraces the messiness of human interpretation of precise, carefully calibrated hazard advisories as wrapped into the inference system. The inference system based on fuzzy logic is accordingly less sensitive to conflict of input information, or small changes in those inputs. Conversely, evaluating a system with information theory \citep{Shannon1948-nc,Weijs2010-tt,Todter2012-ou} with, e.g., cross-entropy loss functions (common in machine learning; often \textit{scoring rules} in meteorology) assumes optimal decision-making associated with its basis in base-2 logarithms \citep{Pierce1980-kt}, and information is hence a pure estimate in distilled units of bits, requiring a weight to represent real-world utility.

\subsubsection{Dual Measures of Possibility and Necessity}
With $\pi$ in hand, and if at least one event is fully possible ($\exists \omega \in \Omega_0 = 1$), we measure necessity (certainty; inevitability) for an event $A$: 

\begin{equation}\label{eq:necessity}
\text{N}(A) = 1 - \Pi(\neg A)
\end{equation}

where the event not occurring, $\neg A$, is defined as

\begin{equation}\label{eq:not_a}
\neg A = 1 - \max(\pi_{\omega \neq A})
\end{equation}

and $\pi_{\omega \neq A}$ is a possibility distribution of all non-$A$ events in $\Omega$, i.e.,

\begin{equation}\label{eq:omega_not_a}
\forall \omega \in \Omega \neq A
\end{equation}

The necessity distribution $\text{N}(A)$ is dependent solely on the possibility distribution $\pi$. We adhere to the three axioms of possibility theory as outlined by \citet{Dubois1988-nh}, but initiate discussion below on \textit{normalisation} of distributions via stretching to ensure $\max_{x} \pi = 1$, fulfilling Axiom 2 below (\textit{something} must happen). This is needed to compute \nec{} but arguably drifts further from physical meaning the more a distribution must be stretched (e.g., $\forall x, x \gtrapprox 0$, implying $\pi^{\prime} \not\approx \pi$ ).

\begin{rhoenv}[
        frametitle=Axioms of Possibility Theory,
        frametitleaboveskip=4pt,
        skipabove=4pt,
        ]\small
\begin{axiomlist}
    \item \textbf{Axiom 1:} $\Pi(A \cup B) = \max(\Pi(A), \Pi(B))$ for events $A, B \in \Omega$. 
    \\ This means that the possibility of the union of two events is determined by the event with the higher possibility, an ``optimistic" property that does not rule out plausible outcomes.

    \item \textbf{Axiom 2:} $\Pi(\emptyset) = 0$ and $\Pi(\Omega) = 1$. 
    \\ This indicates that the impossible event has a possibility of zero, while the certain event (the whole sample space) has a possibility of one.

    \item \textbf{Axiom 3:} If $A \subseteq B$, then $\Pi(A) \leq \Pi(B)$. 
    \\ This states that if event $A$ is a subset of event $B$, the possibility of $A$ cannot exceed that of $B$.
\end{axiomlist}
\end{rhoenv}

Probability theory is rooted in Aristotle's ``exclusion of the middle", or axiom of choice: \textit{something} must happen in the set of possible outcomes, and there are no contradictions, or a probability distribution that sums to greater than unity. Conversely in possibility theory, there are no limits on $\Pi$ values other than constraint in $[0,1]$. \footnote[5]{However, given possibility is based on set membership, should we relax this constraint set by Axiom 2, \citet[][p.16]{Dubois1988-nh} discusses implications for fuzzy sets that are not normalised.} The paradigm deals with conflicting information and transparently indicates ignorance by not strongly activating the ruleset.

\section{Generating Possibilities with Fuzzy Inference Systems: Ozone Prediction Example}\label{sec:fis_example}

\begin{figure}[bth]
    \centering 
    \includegraphics[width=0.99\linewidth]{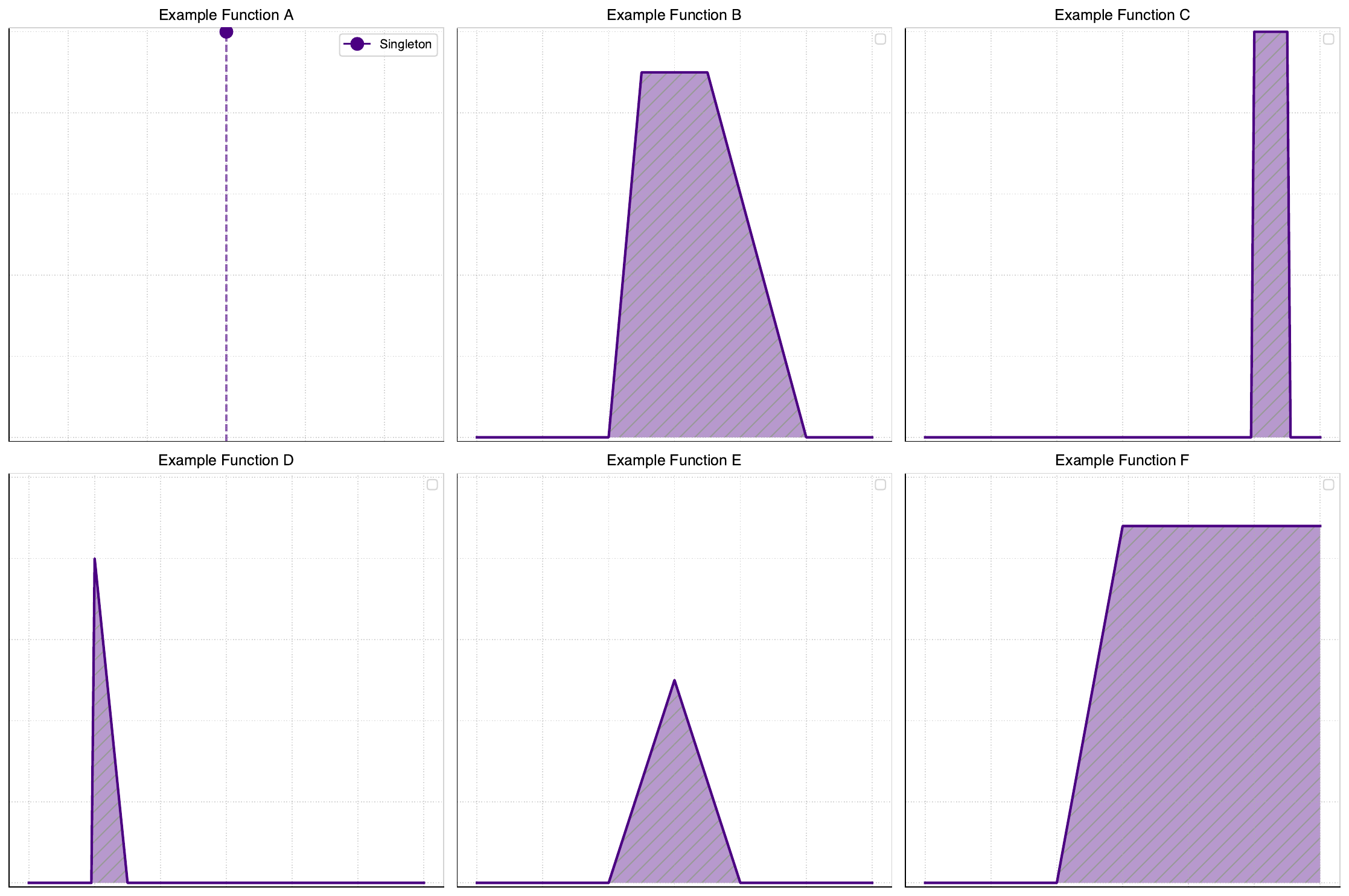}
\caption{A variety of membership function shapes. One-to-one mapping is provided by singletons (panel a), and shapes such as triangles and trapezia (panels b--e)---both symmetric and not---are likewise created by different quintuples shown in Appendix~\ref{appx:mf_creation}. We represent a sigmoid (S-shape; panel f) using a different technique.}\label{fig:multiple_mfs}
\end{figure}

Fuzzy inference systems (FIS) are lightweight models that infer outputs based on input conditions, using a set of fuzzy logic rules. These systems are widely used for control systems \citep{Chevrie1998-go}, predictions \citep{Chang2006-fm,Zounemat-Kermani2008-ul,Camastra2015-by,Saleh2016-wh,Kour2020-rc}, and artificial intelligence applications \citep{Bonissone2010-ik} due to their resilience to uncertainty and flexibility in handling conflicting information \citep{Mamdani1976-pp,Dubois2003-aa}, though fuzzy logic is today considered an elemental form of AI having been superceded by more powerful machine-learning techniques such as artificial neural networks \citep{Shapiro2002-ml, Abraham2005-vp}. However, the theory remains useful for applications beyond its initial use cases and relevant to the present science problem---such as rainfall prediction \citep{Asklany2011-yn}. In this section, we explain key components of a FIS and demonstrate them through a simple ozone-prediction example.

\begin{figure*}[tbh]
    \centering
    \begin{subfigure}[t]{0.3\linewidth}
        \includegraphics[width=\linewidth]{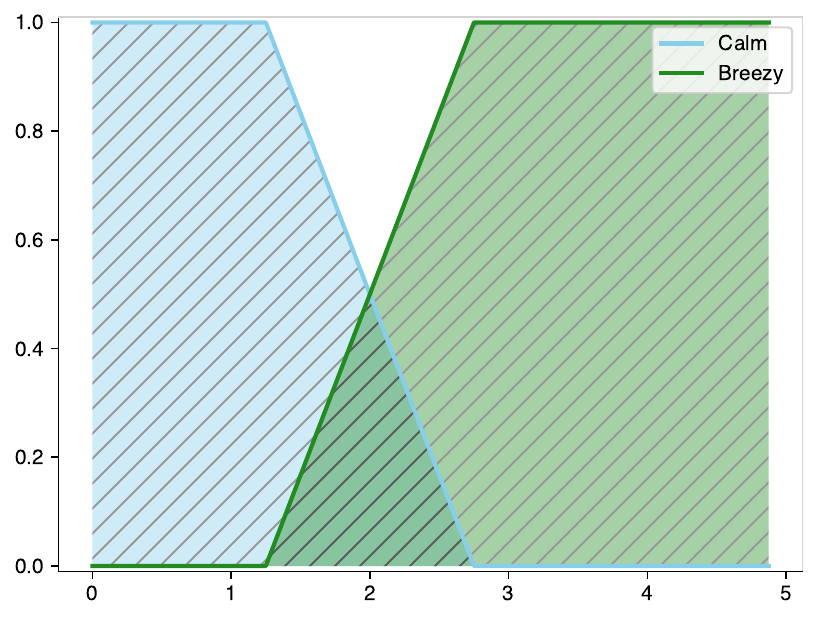}
        \caption{Membership functions of wind speed.}
        \label{fig:mf_wind}
    \end{subfigure}
    \hspace{10pt}
    \begin{subfigure}[t]{0.3\linewidth}
        \includegraphics[width=\linewidth]{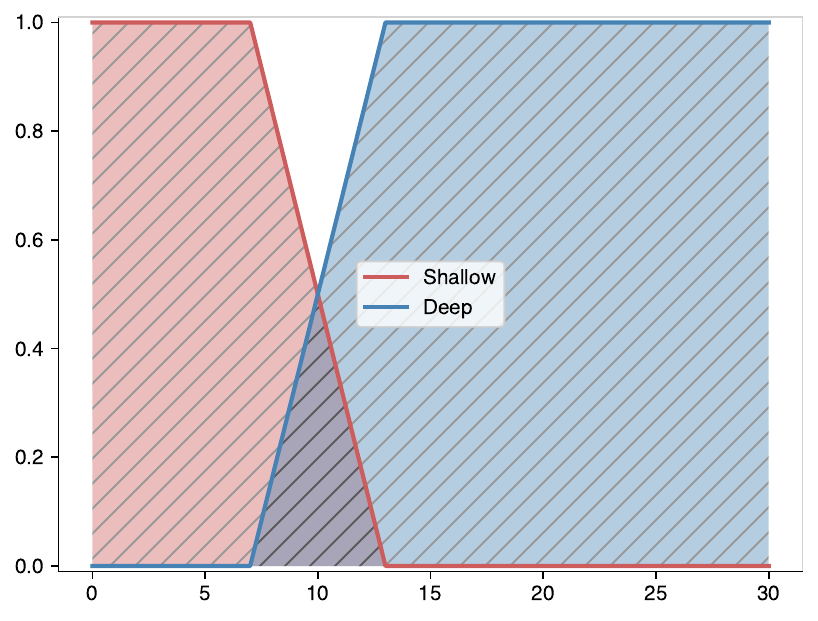}
        \caption{Membership functions of snow depth.}
        \label{fig:mf_snow}
    \end{subfigure}
    \hspace{10pt}
    \begin{subfigure}[t]{0.3\linewidth}
        \includegraphics[width=\linewidth]{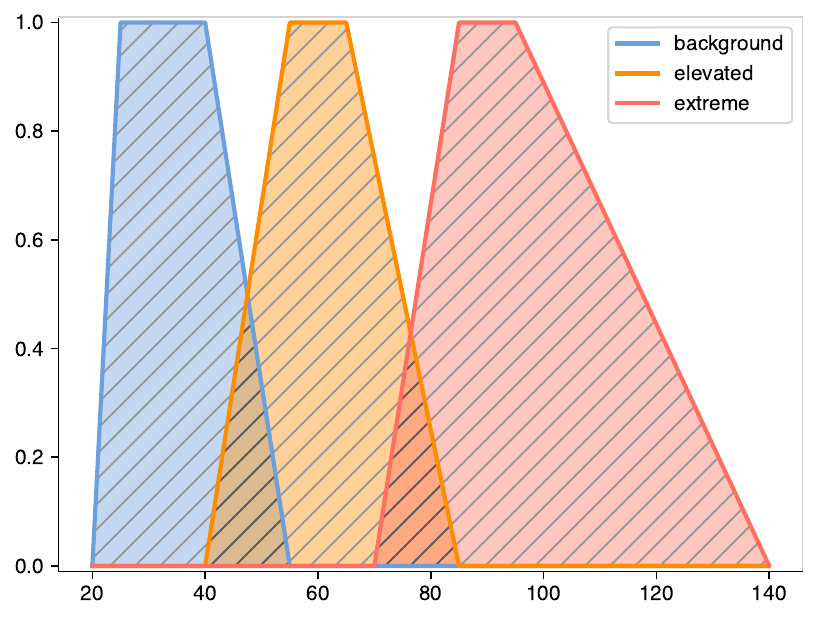}
        \caption{Membership functions of ozone concentration.}
        \label{fig:mf_ozone}
    \end{subfigure}
    \caption{Membership functions for variables in the illustrative system used herein.}
    \label{fig:mfs}
\end{figure*}

\subsection{Prediction problem: Uinta Basin Winter Ozone}
In the Uinta Basin, winter-ozone pollution episodes \citep{Lyman2015-aq,Neemann2015-ws} may occur in the wake of a heavy snowfall if the snowpack can persist, and if surface pressure builds and winds remain calm. This creates cold pools that trap emissions from nearby oil and gas industry. Unlike typical air-quality models that rely heavily on NWP grids, which often struggle with complex terrain and small-scale features that cascade error into air-quality models that depend on atmospheric simulation, the FIS approach leverages rule-based logic informed by domain expertise and historical observations. This approach allows for the system to consider the event-based forecast without the computational burdens of NWP ensembles of high reoslution in time and space. There is no need to ask a model to predict specificity more difficult than necessary, given finite compute power and limited interpretation or development resources. 

\clyfar, the FIS developed for the Basin's winter-ozone physical system \citep{Lawson2024-jb}, ingests meteorological inputs---whether observed or forecast, such as snow depth or wind speed—--as degrees of categories like “sufficient snow” or “calm conditions". The input variables are first transformed to values that represent the Basin en masse to reduce noise from complex orography or unknown aspects of microscale flow. The nature of partial membership means there is lower sensitivity to small changes in input variables \citep{Dubois1988-nh} over, e.g., threshold-based predictions with a tipping point. Assigning memberships to functions encoded human knowledge as machine rulesets: an elementary form of artificial intelligence \citep{Dubois2003-aa}. The formation of rules and representative variables improves trustworthiness in the system due to transparency \citep{Flora2024-de}, and future optimisation with machine-learning \citep[e.g.,][]{Chang2006-fm,Zounemat-Kermani2008-ul} is thus targetted to diagnosed error, more difficult to address in black-boxes such as large neural networks.

\subsection{Membership Functions}\label{sec:mfs}
Membership functions ($\mu$) form the basis of distribution functions $\pi$, whereby each point in this continuous slice of the possible states $x \in \Omega_0$ is mapped to a degree of membership for each category:

\begin{equation}
\mu_x \in [0,1] \qquad \forall\mu_x\in\mu \land x \in \Omega_0
\end{equation}

\begin{figure*}[tbh] 
    \centering
        \begin{subfigure}[t]{0.25\linewidth} 
            \includegraphics[width=\linewidth]{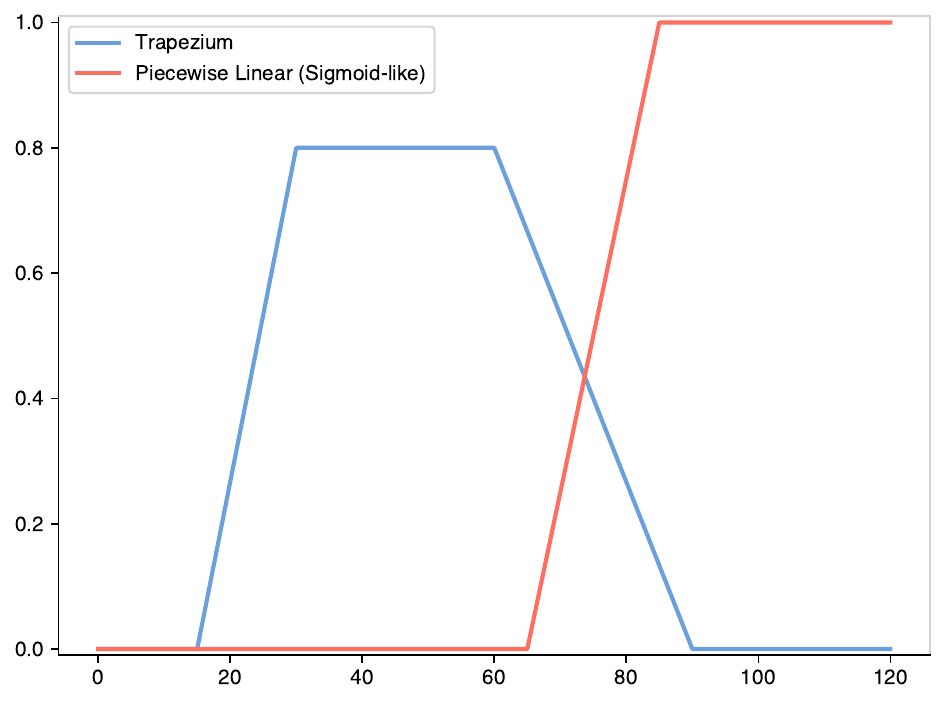}
            \caption{Example of two membership functions of the shapes used herein.}
            \label{fig:mf_a}
        \end{subfigure}
    \hspace{12pt}   
        \begin{subfigure}[t]{0.25\linewidth} 
            \includegraphics[width=\linewidth]{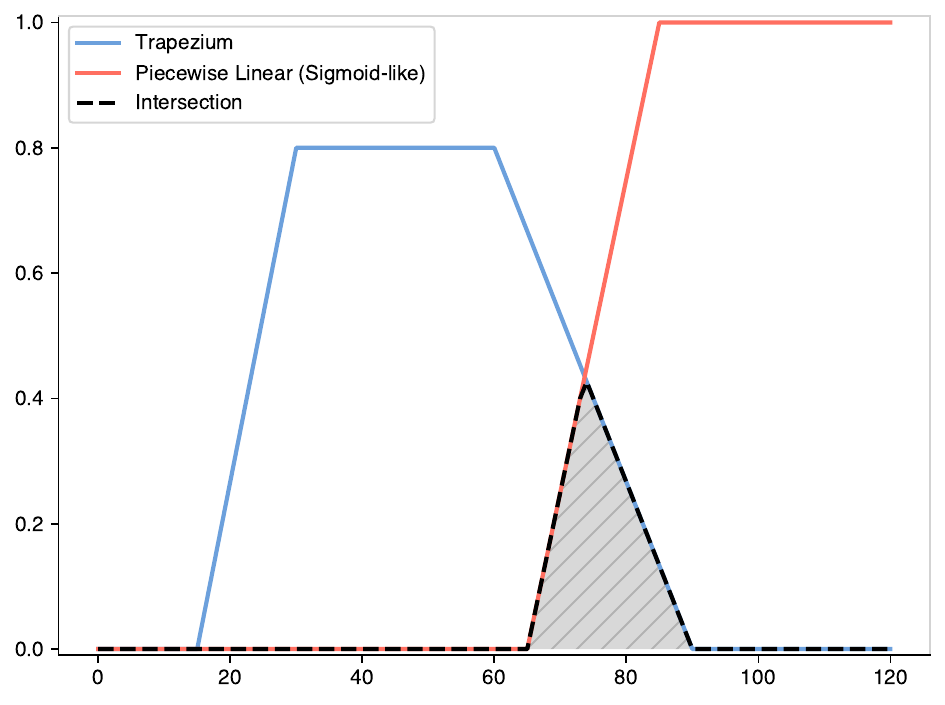}
            \caption{The intersection of two membership functions, implemented as the fuzzy version of the bivalent AND operator (a pessimistic or minimum bound of two possibilities).}
            \label{fig:mf_intersection}
        \end{subfigure}
    \hspace{12pt}   
        \begin{subfigure}[t]{0.25\linewidth} 
            \includegraphics[width=\linewidth]{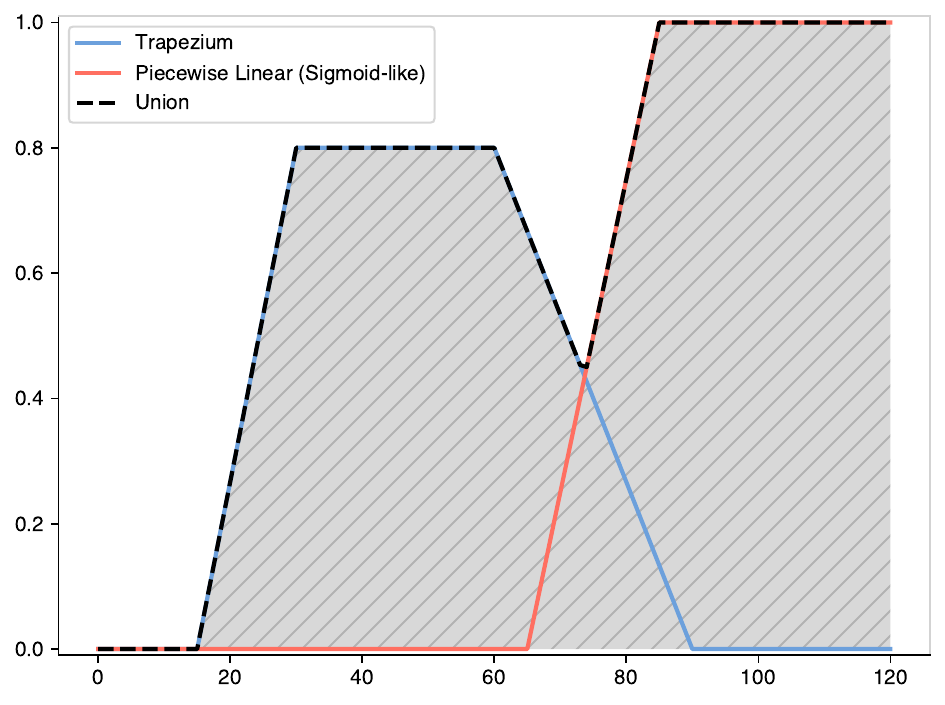}
            \caption{The union of two membership functions, implemented as the fuzzy version of the bivalent OR operator (an optimistic or maximum to combine possibilities).}
            \label{fig:mf_union}
        \end{subfigure}
    \caption{Demonstration of the fuzzy operators joining two membership function shapes used herein: AND (intersection or minimum; panel b) and OR (union or maximum; panel c).}
    \label{fig:operators}
\end{figure*}

Membership functions $\mu$ map input values to degrees of membership in fuzzy sets, enabling the FIS to handle inputs in a nuanced, continuous way. The degree to which an element belongs to a set is defined by this function $\mu$, allowing for partial membership. This is useful for atmospheric predictions, where inputs such as wind speed and snow depth are continuous. We create functions with quintuples that determine shape; we can fine-tune these values with machine-learning to optimise the function shapes (see Future Work). More information about membership functions herein, and rationale behind them, are in Appendix~\ref{appx:mfs}; further graphical presentation of fuzzy inference is given by \citet{Chevrie1998-go}.

\begin{figure*}[tbh] 
    \centering
        \begin{subfigure}[t]{0.3\linewidth} 
            \includegraphics[width=\linewidth]{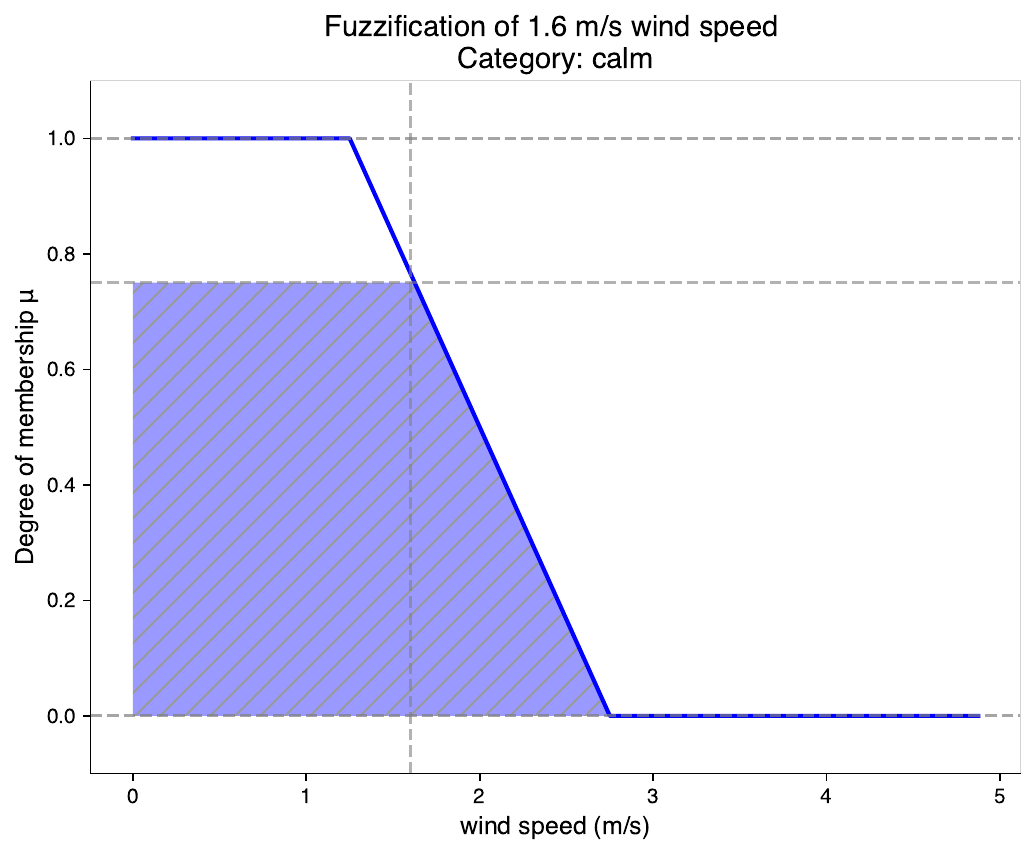}
            \caption{Fuzzification of wind speed as \textit{calm}.}
            \label{fig:fuzz_wind_calm}
        \end{subfigure}
    \hspace{15pt}   
        \begin{subfigure}[t]{0.3\linewidth} 
            \includegraphics[width=\linewidth]{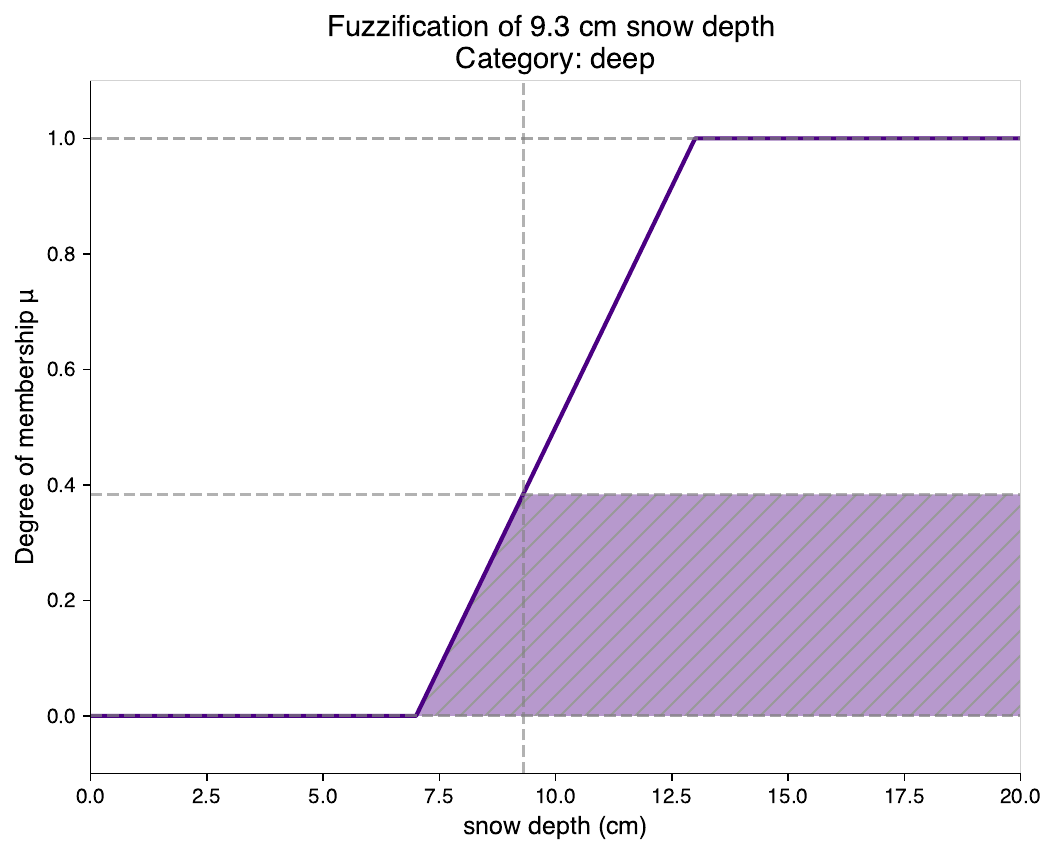}
            \caption{Fuzzification of snow depth as \textit{deep}.}
            \label{fig:fuzz_snow_deep}
        \end{subfigure}
    \hspace{15pt}   
        \begin{subfigure}[t]{0.3\linewidth} 
            \includegraphics[width=\linewidth]{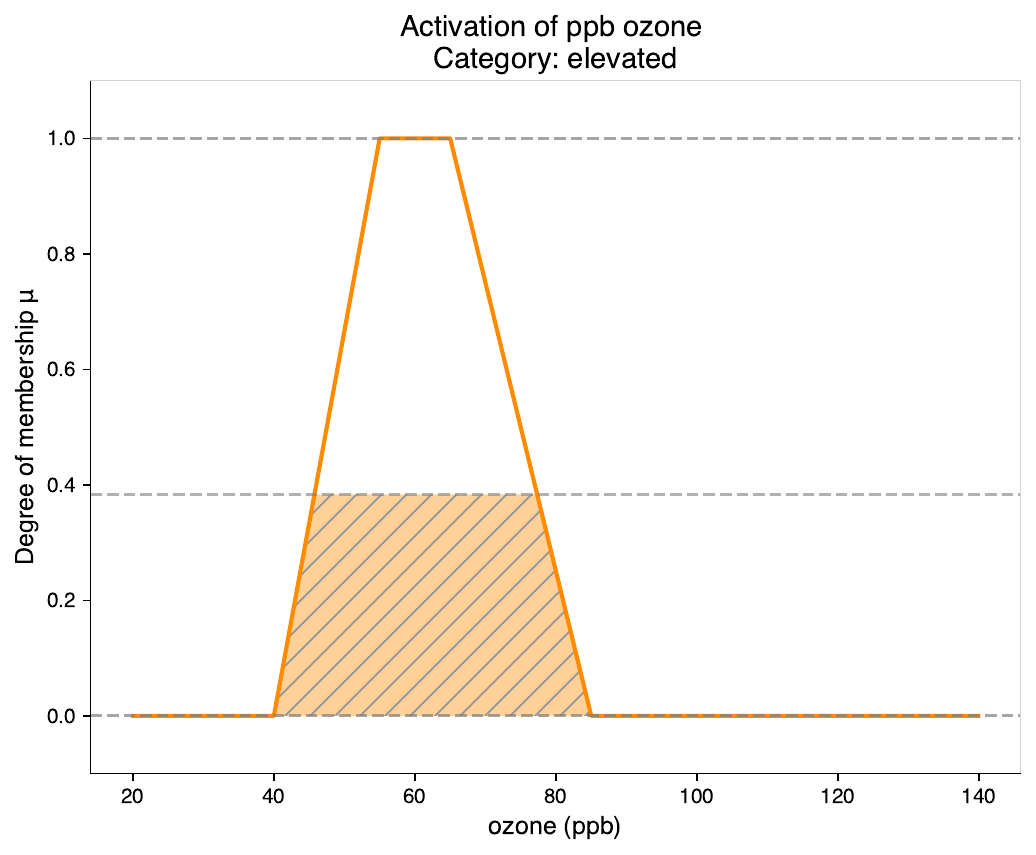}
            \caption{Subsequent activation of \textit{elevated} ozone from Rule 1.}
            \label{fig:activation_ozone_elevated}
        \end{subfigure}
    \caption{The processing of Rule 1: "calm wind and deep snow leads to elevated ozone". Note the activation level in panel (c) is the minimum of activations in (a) and (b).}
    \label{fig:rule1}
\end{figure*}

In this section, we will follow the process of predicting ozone concentration using values of wind speed and snow depth, these values activating a ruleset that produces an ozone value. In this ozone-forecast example, we use two inputs:
\begin{itemize}
    \item \textbf{Wind Speed:} Figure~\ref{fig:fuzz_wind_calm} shows membership of \textit{calm} wind, intuitively assigning higher degrees of membership to lower wind speeds. (Calm conditions promote ozone accumulation through stagnant cold pools.)
    \item \textbf{Snow Depth:} Figure~\ref{fig:fuzz_snow_deep} gives membership for \textit{deep} snow. (Greater accumulation is linked to increased ozone due to albedo precluding warming and extending photon path length for ozone-forming photolysis.
\end{itemize}

We then link the variables with inference rules joined by AND $\land$ operators, again advised by human expertise, explained below.

\subsection{Fuzzy Rules and Rule Activation}\label{sec:rules}
Once inputs are fuzzified through membership functions, the FIS applies fuzzy rules to activate outputs. Rules combine input conditions using logical operators such as AND $\land$ or OR $\lor$, fuzzy cousins of their two-valued equivalent operators. Here, we use solely the AND operator to simplify discussion.

Walking through the inference process, and starting with an observed value for both wind speed and snow depth, we address the first rule (Fig.~\ref{fig:rule1}:

\begin{description}
    \item[\textbf{Rule 1:}] If wind speed is calm AND snow depth is deep, then ozone level is elevated.
\end{description}

\begin{figure*}[tbh] 
    \centering
        \begin{subfigure}[t]{0.3\linewidth} 
            \includegraphics[width=\linewidth]{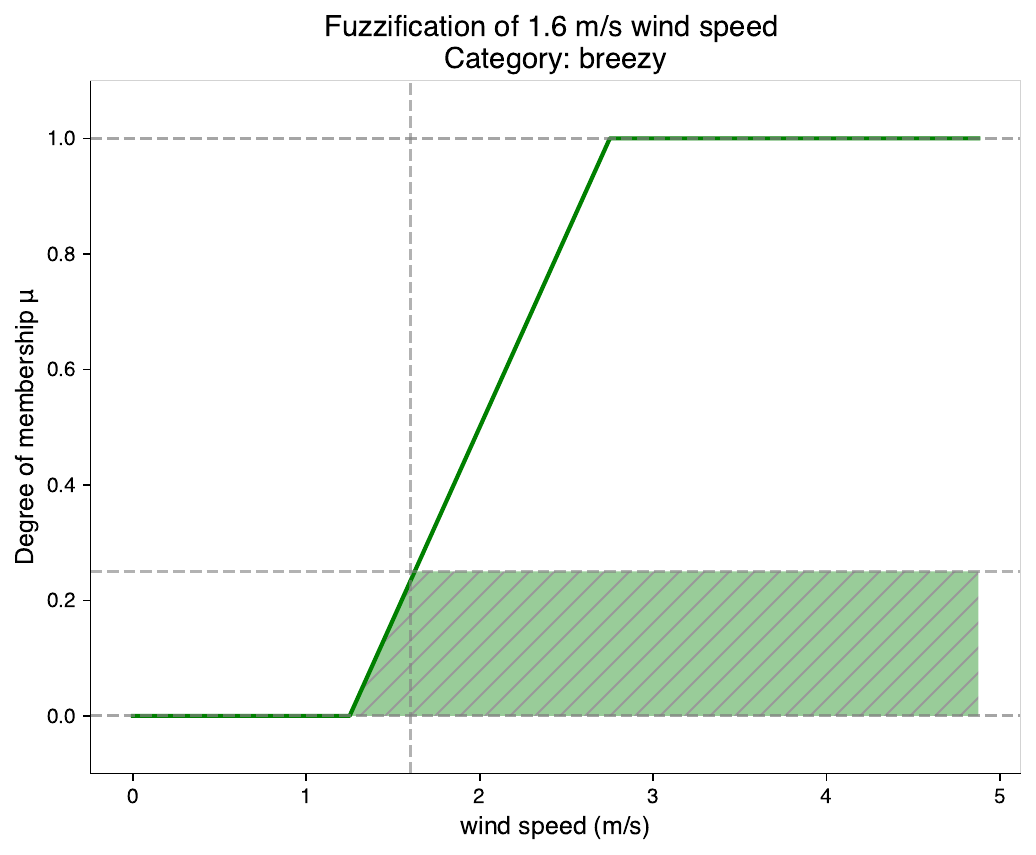}
            \caption{Fuzzification of wind speed as \textit{breezy}.}
            \label{fig:fuzz_wind_breezy}
        \end{subfigure}
    \hspace{15pt}   
        \begin{subfigure}[t]{0.3\linewidth} 
            \includegraphics[width=\linewidth]{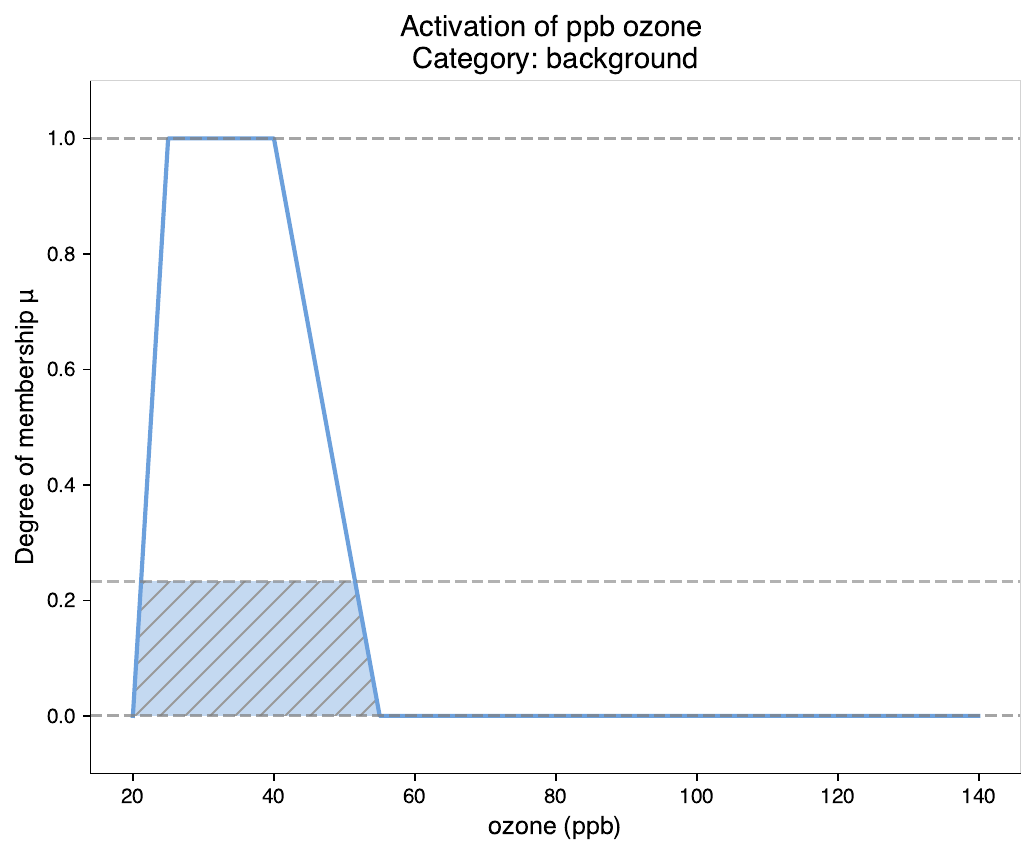}
            \caption{Subsequent activation of \textit{background} ozone concentration from Rule 2.}
            \label{fig:activation_ozone_background}
        \end{subfigure}
    \caption{The processing of Rule 2: "breezy wind implies background ozone". The level of activation is equal to the single input.}
    \label{fig:rule2}
\end{figure*}

\begin{rhoenv}[
        frametitle=Inference method for Rule 1,
        frametitleaboveskip=4pt,
        skipabove=4pt,
        ]\small
    \begin{enumerate}
        \item How calm is the wind? An observed value of 1.6\,$ms^{-1}$ gives a degree of ``calmness" (Fig.~\ref{fig:mf_wind}) to 0.77 or ``substantially deep" (Table~\ref{tab:fuzzy_terms})
        \item How deep is the snow? For a depth of 9.3\,cm this is an activation of 0.38 (``somewhat") deep (Fig.~\ref{fig:mf_snow}).
        \item We combine the AND rules by finding the minimum across all activations, here $\min(0.77,0.38) = 0.38$ (cf. Figs.~\ref{fig:mf_ozone} and \ref{fig:activation_ozone_elevated}).
        \item Hence we can say elevated ozone is \textit{somewhat possible}.
    \end{enumerate}
\end{rhoenv}

\begin{table}[H]
    \centering
    \small
    \setlength{\tabcolsep}{3pt} 
    \caption{Logical operators and associated functions for bivalent logic and fuzzy equivalents, where $A$ and $B$ represent independent events.}
    \label{tab:logical_symbols}
    \begin{tabularx}{\columnwidth}{l>{\centering\arraybackslash}XlX}
        \toprule
        \textbf{Description} & \textbf{Rendered} & \textbf{Bivalent} & \textbf{Fuzzy} \\
        \midrule
        Implication & $\rightarrow$ & & \\
        $A$ AND $B$ & $A \land B$ & minimum & infimum \\
        $A$ OR $B$ & $A \lor B$ & maximum & supremum \\
        NOT $A$ & $\neg A$ & $(1 - A)$ & $(1 - A)$ \\
        \bottomrule
    \end{tabularx}
\end{table}

We add a second rule to define a ruleset governing the relationship between wind speed, snow depth, and ozone levels:
\begin{description}
    \item[\textbf{Rule 2:}] If wind speed is breezy, then ozone level is background, as pollutants are dispersed into the free atmosphere.
\end{description}

The second rule is simpler to compute as it involves one input and one output (Fig.~\ref{fig:rule2}). Note we leave the output category \textit{extreme} untouched by the ruleset at present. This could represent an ignorance of the system, but is here done to focus our discussion.

\begin{figure*}[tbh]
    \centering
    \begin{subfigure}[t]{0.34\linewidth}
        \includegraphics[width=\linewidth]{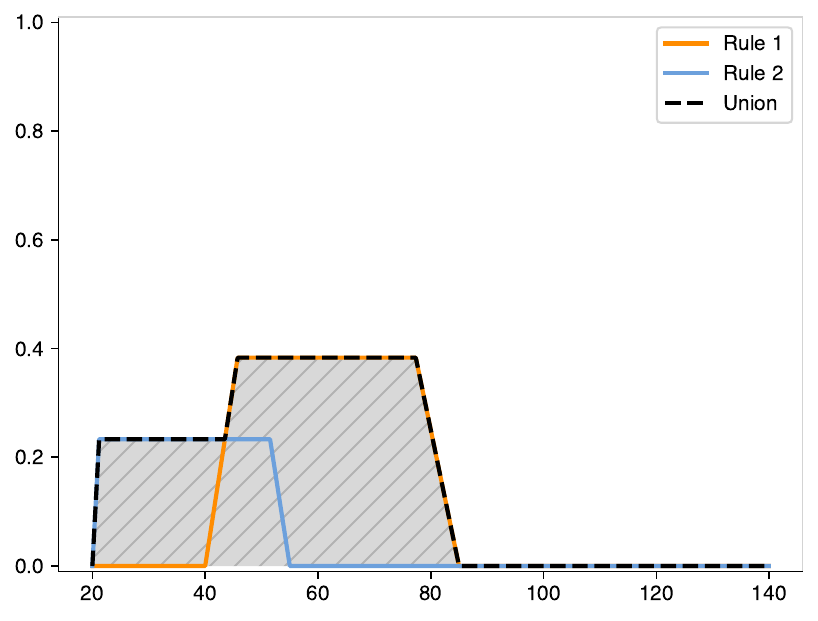}
        \caption{Aggregated activation, representing a possibility distribution $\pi$, formed by the union of distributions formed from rules 1 (Fig.~\ref{fig:activation_ozone_elevated}) and 2 (Fig.~\ref{fig:activation_ozone_background}).}
        \label{fig:aggregated_activation}
    \end{subfigure}
    \hspace{9pt}
    \begin{subfigure}[t]{0.34\linewidth}
        \includegraphics[width=\linewidth]{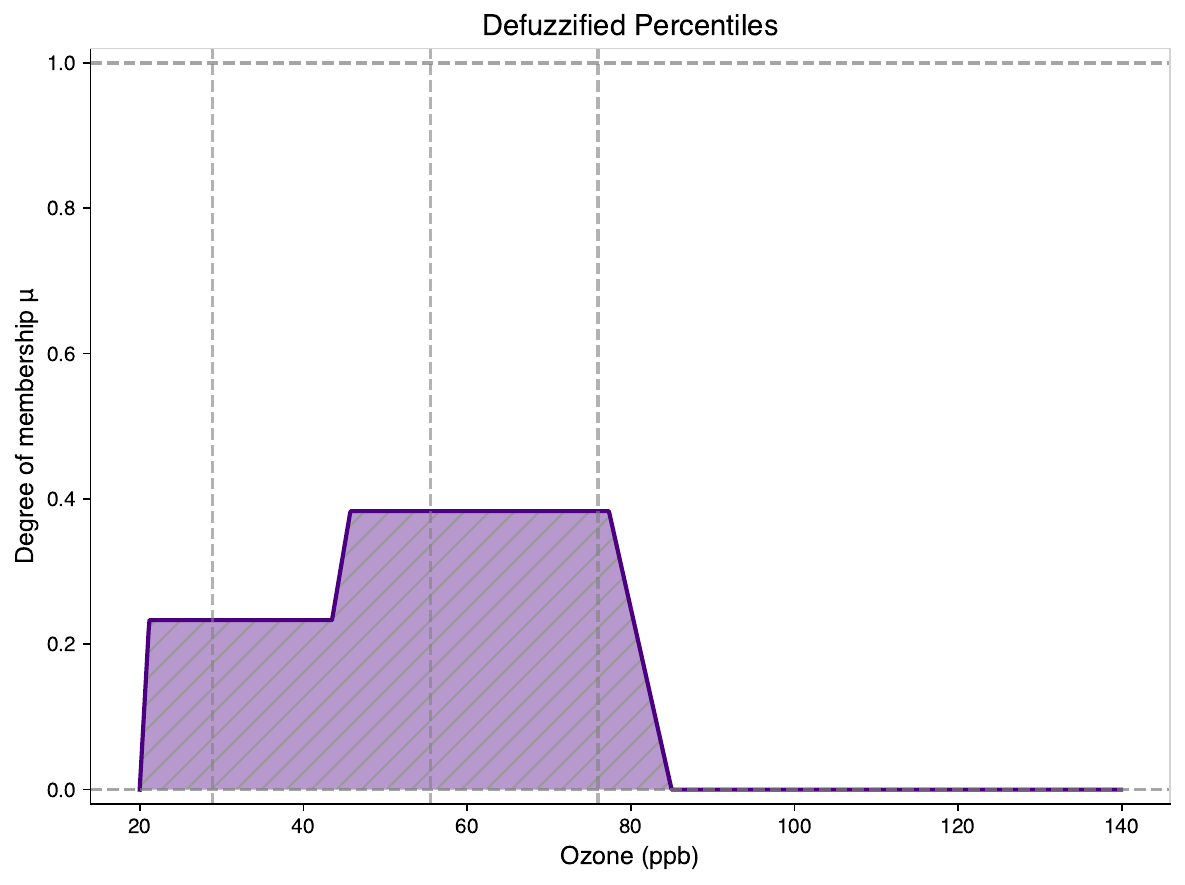}
        \caption{Defuzzification through 10th, 50th, and 90th percentiles of the possibilistic area under the curve.}
        \label{fig:defuzz_pc}
    \end{subfigure}
    \caption{Defuzzification of aggregated activation (a possibility distribution $\pi$).}
    \label{fig:defuzz}
\end{figure*}

\subsection{Aggregation and Defuzzification}\label{sec:agg_defuzz}
Once all rules are evaluated, their output activations are aggregated as in Fig.~\ref{fig:aggregated_activation} with the OR (union or maximum; Fig.~\ref{fig:mf_union}) operator to form a final possibility distribution $\pi$ shown in Fig.~\ref{fig:defuzz_pc}. After aggregation, $\pi$ can be defuzzified to produce a crisp output. In our ozone forecast example, we apply a percentile-based defuzzification method that converts the possibilistic area under the curve $\text{AUC}_{\Pi}$ into specific ozone concentrations (herein, the 10th, 50th, and 90th percentiles), as in Figure~\ref{fig:defuzz}. This range of percentiles may corresponds to reasonable worst-, average-, and best-case scenarios, respectively, if event $A^!$ is undesired. (The code repositories used to calculate these values are linked below in Data Availability.)

The width of this percentile range implies an epistemic uncertainty: a range of possible outcomes or variable values are activated due to imperfect information or suboptimal rulesets. This is an error that can be reduced via machine-learning techniques. Alternatively, we can preserve the curve (or the activation of each set element) to form a sort of category likelihood. The interpretation and visualisation of output that is not defuzzified, i.e., that preserves uncertainty information, follows.

\section{Interpretation of Possibility Distributions}\label{sec:poss_distr}
The area under a possibility distribution $\text{AUC}_{\pi}$ represents a mass of possibility of hazardous event $A^!$. As shown, we can reduce this area to a set of scenarios cutting at various percentiles of area representing potential outcomes. Alternatively, in this section, the author presents ways to preserve the distribution as a measure of uncertainty and magnitude of $\Pi(A^!)$.

\subsection{Residual "Unsure" Possibility Category}\label{sec:unsure}
\begin{figure*}[tb]
    \centering
    \begin{subfigure}[t]{0.33\linewidth}
        \includegraphics[width=\linewidth]{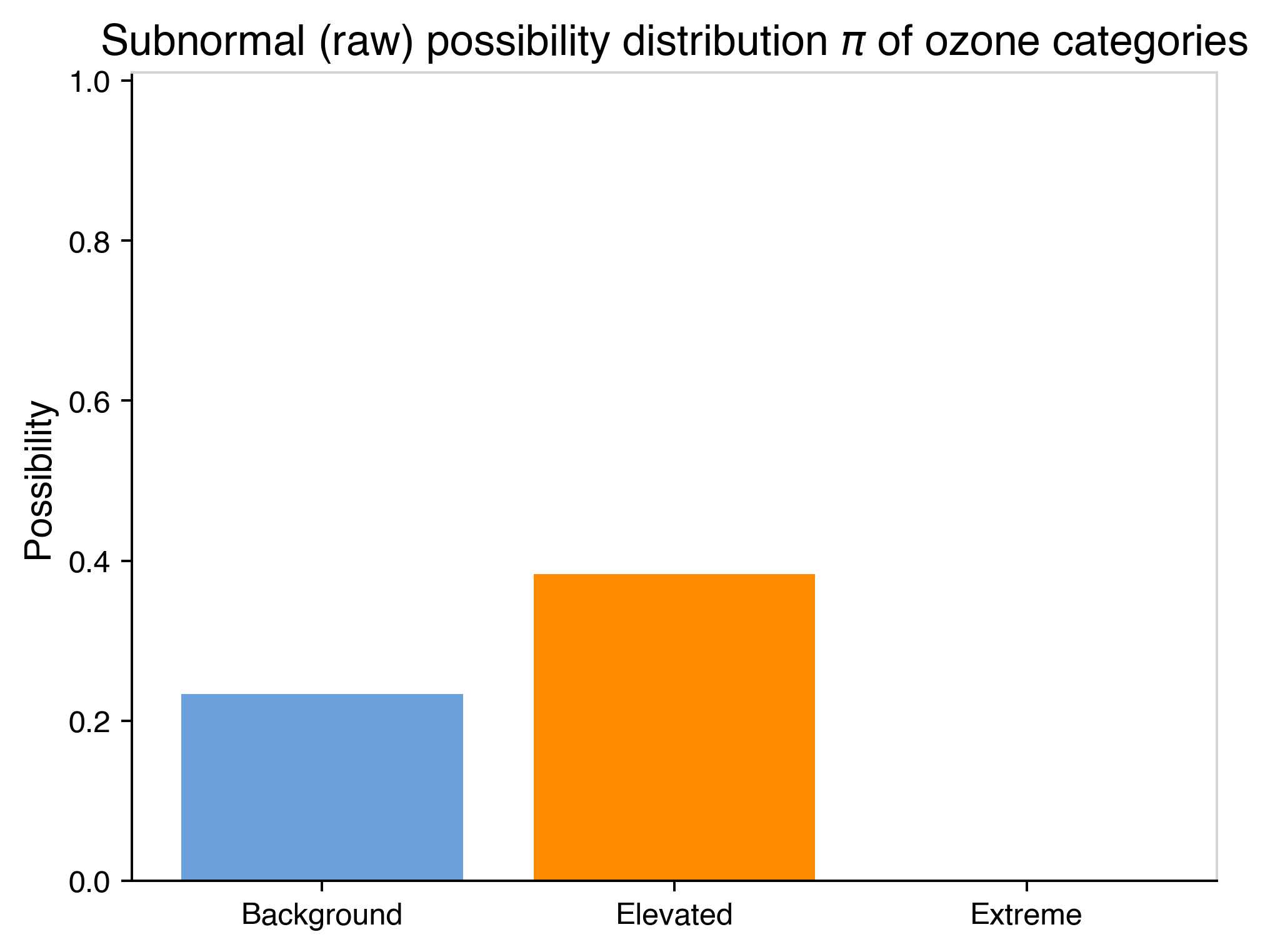}
        \caption{The raw output equivalent to that in Fig.~\ref{fig:aggregated_activation} but presented as bars.}
        \label{fig:poss_raw}
    \end{subfigure}
    \hspace{15pt}
    \begin{subfigure}[t]{0.33\linewidth}
        \includegraphics[width=\linewidth]{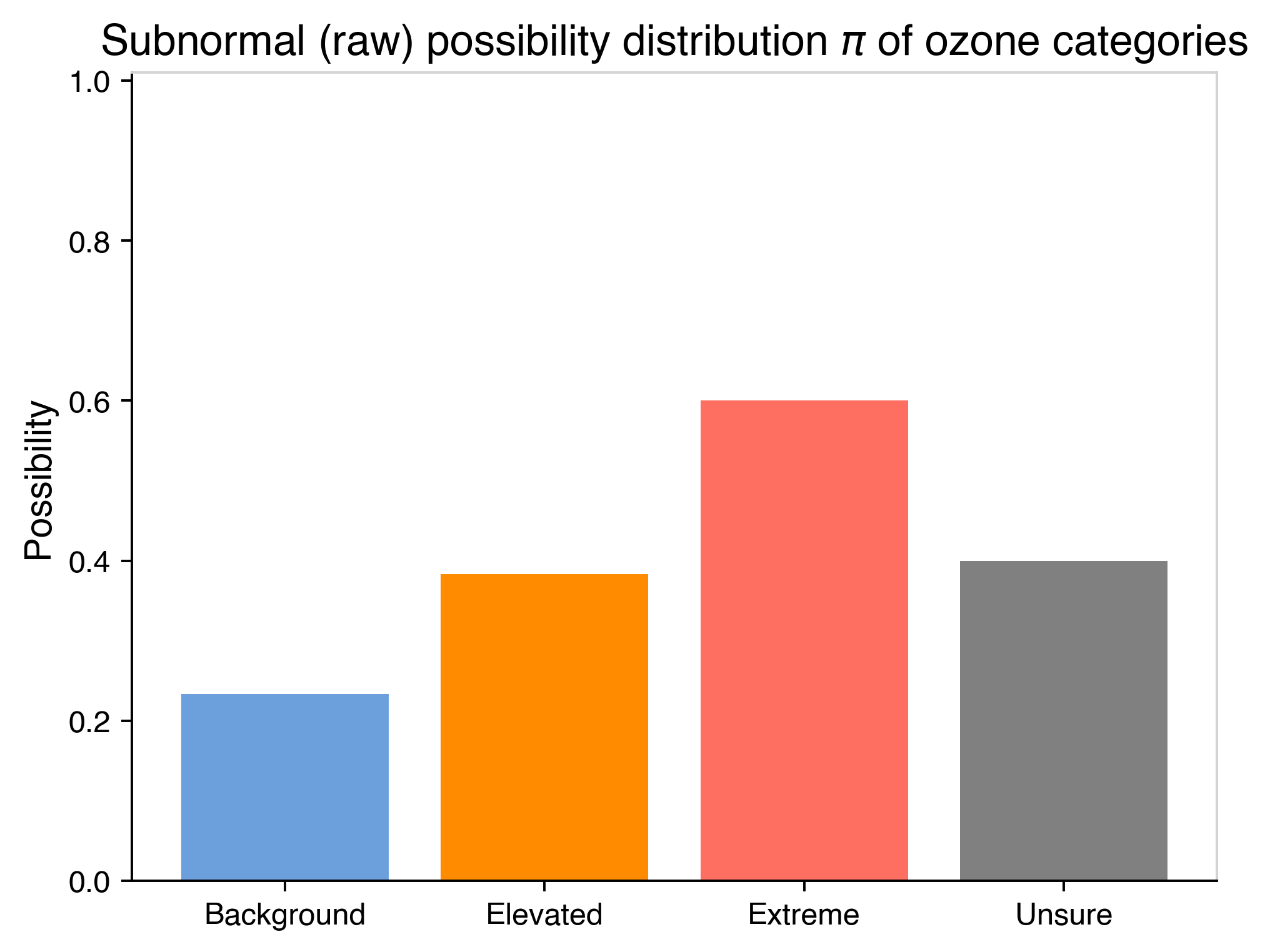}
        \caption{Adding \textit{extreme} and computing \textit{unsure} as the residual possibility unaccounted for in the system.}
        \label{fig:poss_raw_unsure}
    \end{subfigure}\\
    \vspace{10pt}
    \begin{subfigure}[t]{0.33\linewidth}
        \includegraphics[width=\linewidth]{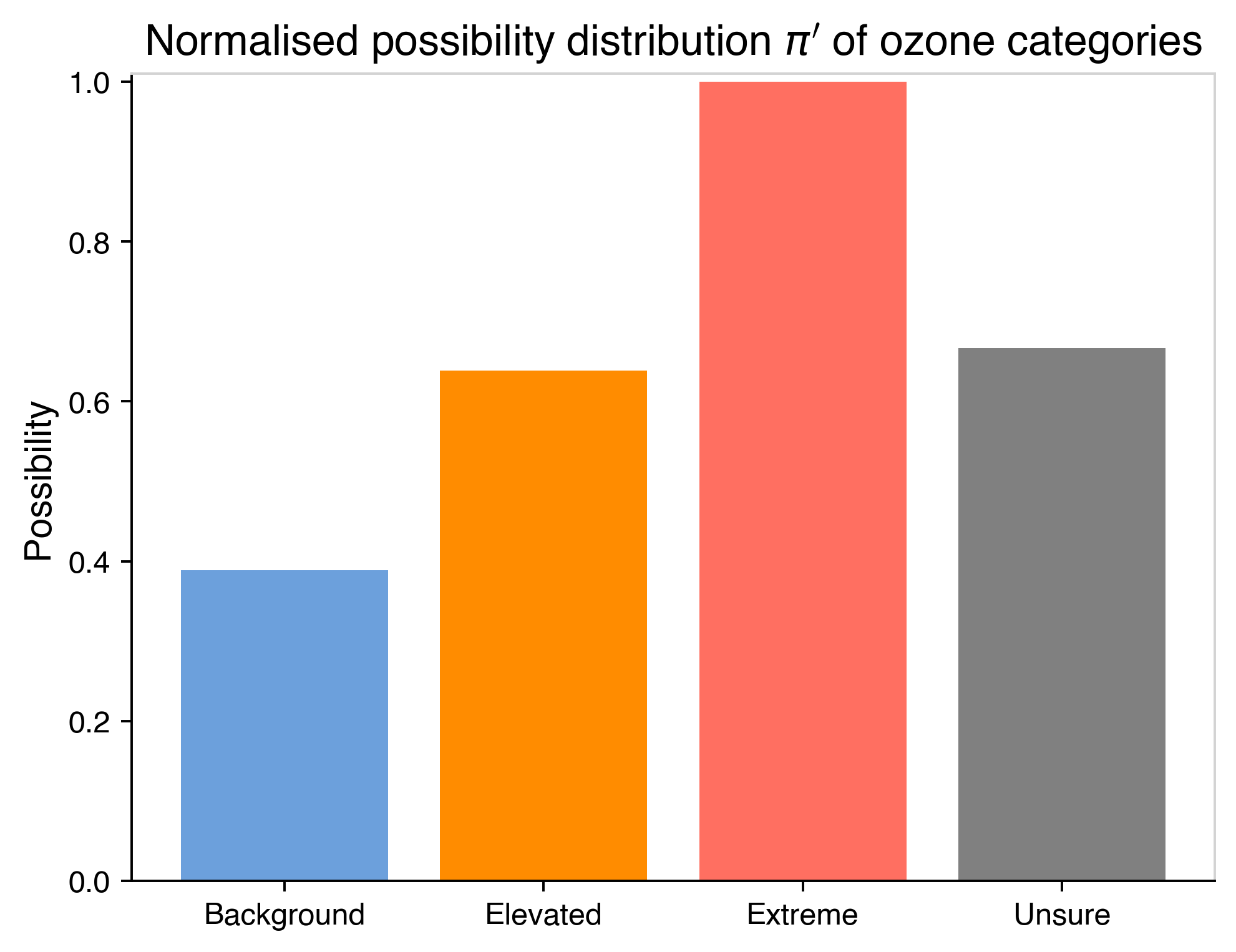}
        \caption{Normalised possibilities (i.e., at least one category is equal to unity).}
        \label{fig:poss_normalized}
    \end{subfigure}
    \hspace{15pt}
    \begin{subfigure}[t]{0.33\linewidth}
        \includegraphics[width=\linewidth]{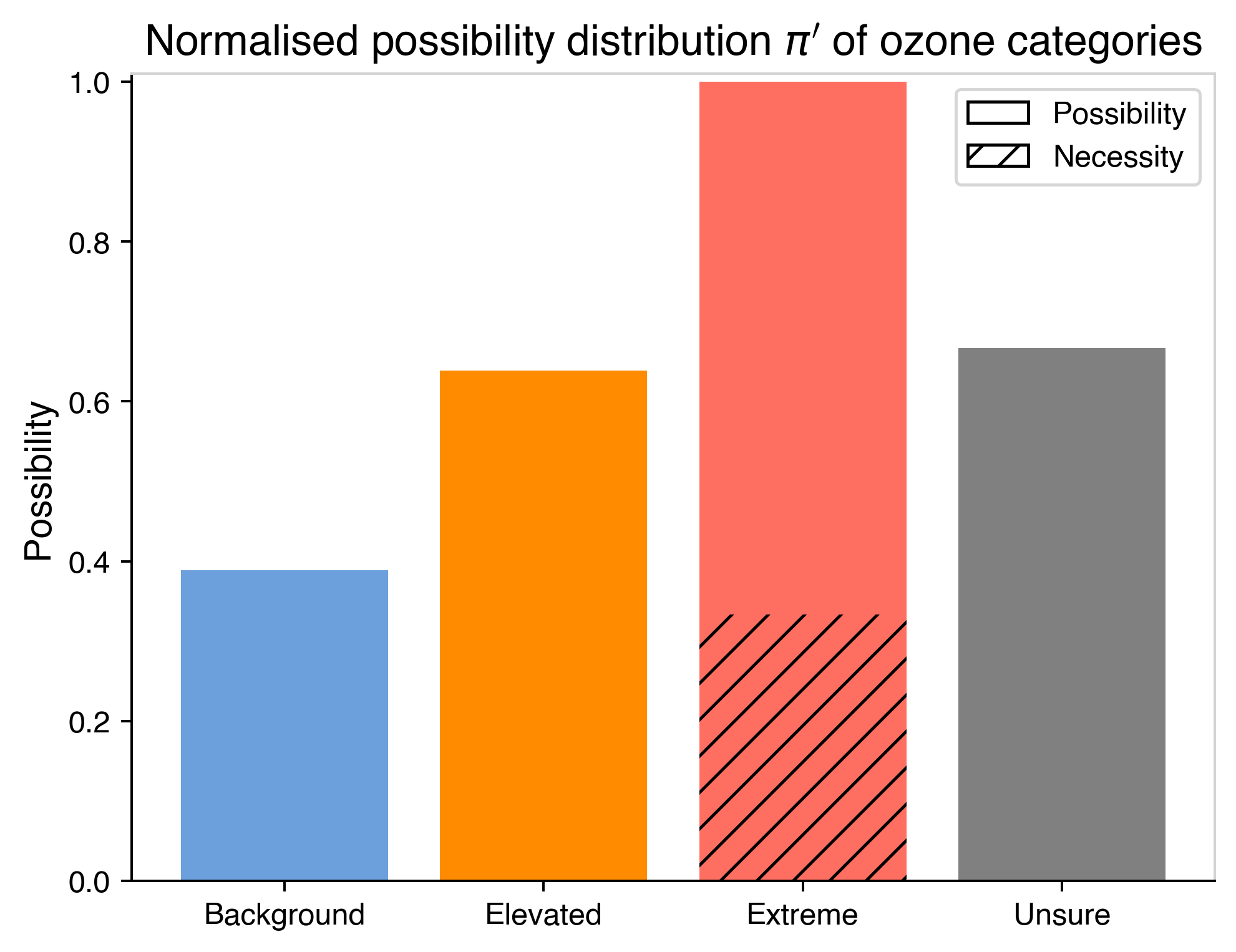}
        \caption{Adding necessity as hatching, representing the need for this event to occur, or an inevitability due to only somewhat plausible alternative outcomes.}
        \label{fig:poss_necessity}
    \end{subfigure}
    \caption{Evolution of Possibility Distributions with Normalization and Necessity.}
    \label{fig:poss_evolution}
\end{figure*}

It is possible for no FIS rule to be fully activated ($\max(\pi) < 1$). The author proposes inclusion of an \textit{unsure} possibility category to address scenarios where the model does not fully cover all possible outcomes (Fig.~\ref{fig:poss_raw}). The value of \textit{unsure} is a measure of both uncertainty types: epistemic uncertainty emerges from a ruleset with insufficient coverage across possible rules and the range of inputs; while aleatoric uncertainty is a natural variability inextricable from the system. If the possibility distribution is not normalised ($\nexists \omega = 1,~\forall \omega \in \Omega_0$), the residual \textit{unsure} category represents the uncertainty in plausibility not accounted for by the existing FIS categories.

\begin{equation}\label{eq:poss_unsure}
    \Pi(\textit{unsure}) = 1 - \max_{x} \pi(x)
\end{equation} 

After adding an arbitrary value for extreme below maximum activation (0.6), the \textit{unsure} category is implicitly activated given other rules are not (see Fig.~\ref{fig:poss_raw_unsure}), hence would be most useful in events not covered by the FIS ruleset, or if there is contradictory information to its configuration (erroneous input data, for instance) that places an upper bound on that rule's activation (due to the AND/minimum operation).

\subsection{Normalisation of the Distribution}\label{sec:normalise}
Normalising the possibility distribution ensures that at least one category equals unity, adhering to Axiom 2 of possibility theory. The advantage of normalisation is access to necessity distribution $\nu$ without violating theoretical axiomata. However, it implies brazenly that the forecasting system covers all possible outcomes. An alternative approach involves creating the \textit{unsure} category for residual uncertainty, which introduces trade-offs.

The normalisation process can be visualised with bar charts, though given our human-derived categories of membership (e.g., shallow or deep snow), these are not independent categories. Nonetheless, they provide a coarse conveyance of the possibility distribution. For example, \citet{Lawson2024-jb} demonstrates bar charts showing aggregated activation representing a possibility distribution $\pi$ from \clyfar{} output.

Figure~\ref{fig:poss_normalized} shows the distribution after normalisation:

\begin{equation}
\pi^{\prime}(x) = \frac{\pi(x)}{\max\left( \pi(x) \right)}, \quad \text{for} \ \pi(x) > 0
\end{equation}

Finally, we compute and plot necessity denoted by hatching on the category or event with $\Pi = 1$: something must be fully possible before it is necessary---or indeed inevitable. Inevitability is borne from a lack of plausibility of categories $\neg A$ that imply $A$ must happen, using Axiom 2: $A$ is highly necessary.

\section{Information Gain and Predictability Horizons}
A continuously optimised model such as \clyfar{} requires something to maximise or minimise, typically the error between a forecast value and the observed ``truth". It is often clear how this is performed for crisp, sharp values (e.g., 70\,ppb), typically minimising a perceived distance from optimal (hedging). However, accounting for uncertainty requires assessment of multiple events to gather sufficient sampling. Development in theories of both \textit{information} \citep{Hartley1928-uf,Shannon1948-nc} and \textit{chaos} \citep{Lorenz1963-zy,May1976-cy} were grounded in probability. In our case, we can state our goal as \textbf{reducing surprise about future events}, and surprise is equal to information content (entropy; \citealt{Cover2012-di}). 

While \textit{information} is rigorously measured \citep{Cover2012-di}, it is not immediately clear how the information or utility content of a possibilistic forecast would be gauged: calibration is typically assessed in probabilistic models by equating the concept of frequency with predicted percentage (frequentist versus Bayesian statistical paradigms are discussed in \cite{Fornacon-Wood2022-lf}, amongst others). As a starting point, however, the author asserts it is intuitive that more information is gained when the possibility distribution is tall before normalization (i.e., $\max\pi \approx \max\pi^{\prime} = 1$), has a small range in the x-direction (i.e., high confidence, or a lower ``statistical entropy"), exhibits high necessity (i.e., low possibility for other outcomes, hence an inevitability for the event of interest), and when an event is rare or difficult to predict \textit{a priori}. These characteristics convey confidence in specificity, or high necessity for an event to happen.

\subsection{Reducing Surprise in Chaotic Systems}
First, predictability herein is defined as an intrinsic property of an event within a chaotic, complex system such as the climate system. It is inherent to the flow or state and \textit{not} measured by prediction skill alone. Predictability is often limited by sensitivity to initial conditions \citep{Palmer2014-cv}. Possibility theory can extend the information gain horizon by trading specificity (probabilities) for possibilities with associated certainty measures. This allows for a broader representation of potential outcomes without requiring extensive historical data.

We identify locations of low predictability in time and space:
\begin{enumerate}
    \item At the temporal limit relative to the weather phenomenon: \citet{Lorenz1963-zy,Lorenz2004-qw} estimated 10--14 days on the continental scale),
    \item On a cusp or tipping point between two solutions, either in models (5 in 10 members do simulate $A^!$) and/or in reality (future state is close to a bifurcation),
    \item Weaknesses in model or configuration (\textit{black swan} risk),
    \item Inherently low predictability \citep{Palmer2005-xv} perhaps manifest through combinations of the other events by appearing as a consistently low-probability outcome in multiple models with otherwise good fidelity.
\end{enumerate}

In recapitulation, deployment of possibility reduces surprise for risk-averse users by extending forecasts as far into the future as possible or as precisely as possible (i.e., with high confidence). This is particularly useful for complex problems requiring finer geographical specificity for forecast validity. Predicting a possibility is inherently easier than predicting the exact probability of an event occurring, especially for rare events where sample sizes are insufficient for reliable frequentist-style probabilities. 

\section{Summary and Future Work}\label{sec:summary}
\rhostart{T}his study addresses a common obstacle in frequentist and Bayesian probability paradigms: estimating frequency of a rare event with a small sample size. There are scenarios where an even confidence between two future states are supported by data (number of simulations with that solution) or not (an unprecedented event), yet probability theory cannot distinguish which scenario is more supported by data. We propose possibility theory as another tool complementing traditional probability for expressing forecast uncertainty, particularly in low-predictability scenarios where risk-averse users benefit most from information at the predictability horizon. Herein, we generated forecast outputs as degrees of plausibility (\textit{possibility}) and certainty (\textit{necessity}) using a fuzzy-logic inference system \clyfar{} used for predicting ozone concentrations. We demonstrate how input variables like wind speed and snow depth can produce possibility distributions of the forecast variable (ozone) by activating ruleset formed with linguistic terms (e.g., \textit{deep} snow). This grounds lexical abstractions in real-world conditions and \textit{vice versa}: natural language chosen can be tailored in complexity for the researcher or end-user. This possibility framework prioritises early-stage detection of uncertain events, trading absolute specificity for broader coverage of potential outcomes—a trade-off that holds distinct value near predictability limits.

Moreover, by differentiating between aleatoric (inherent randomness) and epistemic (knowledge-based) uncertainty, possibility theory allows for a more nuanced representation of forecast confidence. The introduction of an \textit{unsure} category for the output variable further captures cases where traditional rulesets may lack full coverage of the ``universe of possibilities", ensuring that uncertainty is transparently communicated rather than concealed by overconfident predictions. While additional work remains to integrate possibility theory into accessible language for operational forecasting and public engagement---and a need remains to develop an evaluation framework (Appendix~\ref{appx:info_gain})---this study reveals its utility for addressing the dual uncertainties when predicting states of complex systems like the climate (or indeed a mountain-basin cold pool). 

\subsection{Future Work}\label{sec:future}
Future efforts should first focus on optimising fuzzy-logic membership functions through machine learning to improve rule activation and optimisation (loss minimisation), though how this is measured in a possibilistic paradigm is yet to be determined. Methods such as genetic algorithms \citep{Roebber2015-vv,Roebber2015-ps} and Grey Wolf \citep{Zhang2024-kf} demonstrated on similar problems can discover new rulesets that contribute value to the prediction. This evolution maintains an understandable, trustworthy AI system only as complex as it must be to deliver useful information, rather than reaching straight for a \textit{black-box}. The system need not be fully understandable by the user, but its answers should be trustworthy---further discussion is found in \citet{Flora2024-de}.

Planned projects include translating complex guidance into understandable formats and integrating large-language models to enhance communication of forecast uncertainty, including visualisations \citep[e.g.,][]{Lawson2024-pa}. There is a need to communicate possibilities in natural language suitable for a variety of users, particularly in the author's research setting at the nexus of many varied stakeholders \citep{Lyman2023-sa}.

\section*{Acknowledgments}
The author thanks Seth Lyman, Austin Coleman, James Correia, Jr., and Corey Potvin for stimulating conversations on the complexity of communicating uncertainty. This work is funded by \textit{Uintah County Special Service District 1 and the Utah Legislature}.

\section*{Use of Generative AI}
The author used OpenAI's large-language models \texttt{GPT-4o} and \texttt{GPT-o1} to link concepts and brainstorm operational research avenues. No text generated by AI was used verbatim in this manuscript. All factual statements given by AI were cross-checked against other literature. 

\section{Data Availability}\label{sec:data_avail}
Code used to generate figures herein is contained within the repository for the full \clyfar{} forecast system at \url{www.github.com/bingham-research-center}.

\section{Glossary}\label{sec:glossary}
See Table~\ref{tab:glossary}.

\begin{table}[h!]
    \RaggedRight
    \caption{Glossary of terms herein}
    \label{tab:glossary}
    \setlength{\tabcolsep}{0.8pt} 
    \renewcommand{\arraystretch}{1.01} 
    \small
    \begin{tabular}{@{}p{0.36\columnwidth}p{0.57\columnwidth}@{}}
        \toprule
        \textbf{Term} & \textbf{Definition} \\ 
        \midrule
        Universe of Discourse ($\Omega$) & Complete set of all possible events under consideration. \\
        Specific event ($A$) or high-impact event ($A^!$) & Specific event in total set $\omega \in \Omega$.\\
        Possibility Distribution ($\pi$) & Function assigning a possibility value to each event in subset of possible states $\Omega_0 \subset \Omega$. \\
        Normalised poss. distr. ($\pi^{\prime}$) & Stretched possibility distribution ensuring $\max_{x}\pi = 1$. \\
        Necessity Distribution ($\nu$) & Function derived from $\pi$ assigning a necessity value to each element in same event subset $\Omega_0$. \\
        Possibility ($\Pi(A)$) & A measure indicating plausibility or support in climatology that event $A$ may well occur. \\
        Necessity ($\text{N}(A)$) & A measure indicating the certainty or inevitability that event $A$ must occur. \\
        Fuzzy Inference System (FIS) & A system deploying fuzzy-logic rules to infer possibility distributions from input data. \\
        Defuzzification & The process of converting a fuzzy set (distribution $\pi$) into a crisp value. \\
        Membership Function ($\mu$) & Function similar to $\pi$ whereby each point in input space is mapped to a degree of membership for each category. \\
        \clyfar{} & Our ozone-prediction FIS, named for the Welsh word for "clever", or \emph{Computational Logic Yielding Forecasts for Atmospheric Research}. \\
        \bottomrule
    \end{tabular}
    \vspace{0.5em}
\end{table}


\section{Appendix}\label{sec:appx}

\subsection{Creating membership functions}\label{appx:mfs}

\subsubsection*{Membership function creation}\label{appx:mf_creation}
We use two membership function shapes herein: trapezoidal and piecewise-linear (to create sigmoid-like curves), as shown in Figure~\ref{fig:mfs}. There are benefits to using piecewise linear functions, described by \citet{Dubois1988-nh} and mostly outside the scope of the present study (e.g., regarding calculus of distributions). Herein, we choose piecewise linear membership functions primarily to (1) simplify calculations of area under the curve, and (2) increase possibility mass at high $\Pi$ and $\mu$ values instead of peaked Gaussian membership functions, a decision informed by results following further testing of a \clyfar{} prototype \citep{Lawson2024-jb}. These functions define fuzzy sets for the meteorological variables we use in the FIS.

\begin{table}[htbp]
\centering
\small 
\setlength{\tabcolsep}{4pt} 
\caption{Parameters for membership-function shapes. The five quantities for trapezia (the ``quintuple") are described in text. Piecewise-linear approximations of sigmoids are processed differently below.}
\begin{tabular}{@{}lccccc@{}}
\toprule
\textbf{Source} & $m_{\text{lower}}$ & $m_{\text{upper}}$ & $\alpha$ & $\beta$ & $h$ \\ 
\midrule
\multicolumn{6}{c}{\textbf{Trapezoidal Membership Functions}} \\ \midrule
\textbf{A} & 100 & 100 & 0 & 0 & 1 \\
\textbf{B} & 50 & 70 & 10 & 30 & 0.9 \\
\textbf{C} & 100 & 110 & 0 & 0 & 1 \\
\textbf{D} & 20 & 20 & 0 & 10 & 0.8 \\
\textbf{E} & 60 & 60 & 20 & 20 & 0.5 \\ 
\midrule
\multicolumn{6}{c}{\textbf{Piecewise Linear Sigmoid Function}} \\ \midrule
\textbf{F} & \multicolumn{2}{c}{\textbf{Midpoint}} & \textbf{Width} & \textbf{Direction} & \textbf{Height} \\ 
\textbf{} & \multicolumn{2}{c}{50} & 20 & Increasing & 0.88 \\ 
\bottomrule
\end{tabular}\label{tab:mf_params}
\end{table}

\subsubsection*{Piecewise Linear Sigmoid Function}

The piecewise linear sigmoid function approximates the shape of a sigmoid curve using straight lines. The function is defined by the midpoint, width, height, and direction (either increasing or decreasing). The general form of the piecewise linear sigmoid function is:

$
y(x) = 
\begin{cases} 
  0 & \text{if } x < \text{left}, \\
  \frac{h}{w} \cdot (x - \text{left}) & \text{if } \text{left} \leq x \leq \text{right}, \\
  h & \text{if } x > \text{right}.
\end{cases}
$

Where:
\begin{itemize}
    \item $h$ is the height of the function, left as variable for capturing uncertainty explicitly,
    \item $w = \text{right} - \text{left}$ is the width of the sigmoid region,
    \item $\text{left} = \text{midpoint} - \frac{w}{2}$ is the left boundary,
    \item $\text{right} = \text{midpoint} + \frac{w}{2}$ is the right boundary.
\end{itemize}

If the sigmoid is increasing, the function rises from 0 to the maximum height $h$ within the interval $[\text{left}, \text{right}]$. If the sigmoid is decreasing, the function follows the same structure but decreases from $h$ to 0 over the interval.

\subsubsection*{Trapezoidal Membership Function}

The trapezoidal membership function is based on the quintuple-based definition from \citet{Dubois1988-nh}. This quintuple defines a lower boundary ($m_{\text{lower}}$), an upper boundary ($m_{\text{upper}}$), and slopes determined by $\alpha$ (left slope) and $\beta$ (right slope). The height of the trapezoid is denoted by $h$, whose value is permissible within $[0,1]$, and where non-unity values indicates less-than-full membership of a given category or set). The general form of the trapezoidal function is:

\[
y(x) = 
\begin{cases} 
  0 & \text{if } x < m_{\text{lower}} - \alpha, \\
  \frac{h}{\alpha} \cdot (x - (m_{\text{lower}} - \alpha)) & \text{if } m_{\text{lower}} - \alpha \leq x < m_{\text{lower}}, \\
  h & \text{if } m_{\text{lower}} \leq x \leq m_{\text{upper}}, \\
  \frac{h}{\beta} \cdot (m_{\text{upper}} + \beta - x) & \text{if } m_{\text{upper}} < x \leq m_{\text{upper}} + \beta, \\
  0 & \text{if } x > m_{\text{upper}} + \beta.
\end{cases}
\]

Where:
\begin{itemize}
    \item $\alpha$ is the distance from $m_{\text{lower}}$ to the start of the trapezoid's slope,
    \item $\beta$ is the distance from $m_{\text{upper}}$ to the end of the trapezoid's slope,
    \item $h$ is the height of the trapezoid.
\end{itemize}

The function has three key parts:
\begin{itemize}
    \item A rising slope from $0$ to the height $h$ over the interval $[m_{\text{lower}} - \alpha, m_{\text{lower}}]$,
    \item A flat top between $m_{\text{lower}}$ and $m_{\text{upper}}$ with constant height $h$,
    \item A falling slope from $h$ back to $0$ over the interval $[m_{\text{upper}}, m_{\text{upper}} + \beta]$.
\end{itemize}

This formalisation will enable understanding fine-tuning of $\mu$ shapes with machine learning as in \citet{Kar2014-nr} and studies discussed therein.

\subsection{Valid and Invalid $\Pi$--$\textnormal{N}$ Configurations}\label{appx:valid_invalid}

For further interpretation, we summarise scenarios with different uncertainty characteristics as measured by dual components $(\Pi, \textnormal{N})$.

\subsubsection*{Valid Solutions}

\begin{table}[h!]
    \centering
    \caption{Valid configurations of possibility and necessity}
    \label{tab:valid_solutions}
    \begin{tabular}{@{}ccc@{}}
        \toprule
        \textbf{Case} & \textbf{Possibility} $\mathbf{\Pi(A)}$ & \textbf{Necessity} $\mathbf{\text{N}(A)}$ \\ 
        \midrule
        1 & \textcolor{possibilityColor}{$1$} & \textcolor{necessityColor}{$0$} \\
        2 & \textcolor{possibilityColor}{$1$} & $0 < \text{N}(A) < 1$ \\
        3 & \textcolor{possibilityColor}{$1$} & \textcolor{necessityColor}{$1$} \\
        4 & $\Pi(A) < 1$ & \textcolor{necessityColor}{$0$} \\
        \bottomrule
    \end{tabular}\label{tab:valid_poss_nec}
    \vspace{0.5em} 
\end{table}

\begin{rhoenv}[
        frametitle=Cases of valid $\Pi$--N dual measures,
        frametitleaboveskip=4pt,
        skipabove=4pt,
        ]\small
\noindent
\begin{description}[leftmargin=!, labelwidth=\widthof{\bfseries Case 4:}]
    \item[Case 1:] $\Pi(A) = 1$ and $\text{N}(A) = 0$. Here, $A$ is fully possible but not necessary.
    \item[Case 2:] $\Pi(A) = 1$ and $0 < \text{N}(A) < 1$. In this scenario, $A$ is fully possible with some degree of necessity.
    \item[Case 3:] $\Pi(A) = 1$ and $\text{N}(A) = 1$. Thus, $A$ is fully possible and necessary.
    \item[Case 4:] $\Pi(A) < 1$ and $\text{N}(A) = 0$. Therefore, $A$ is not fully possible and not necessary.
\end{description}
\end{rhoenv}

\subsubsection*{Invalid Solutions}

There is one general case of invalidity, pertaining to the logical fact something must be fully possible before it must be inevitable:

\begin{table}[h!]
    \centering
    \caption{Invalid configuration of possibility and necessity}
    \label{tab:invalid_solutions}
    \begin{tabular}{@{}ccc@{}}
        \toprule
        \textbf{Case} & \textbf{Possibility} $\mathbf{\Pi(A)}$ & \textbf{Necessity} $\mathbf{\text{N}(A)}$ \\ 
        \midrule
        1 & $\Pi(A) < 1$ & $\text{N}(A) > 0$ \\
        \bottomrule
    \end{tabular}\label{tab:invalid_poss_nec}
    \vspace{0.5em}
\end{table}

\begin{rhoenv}[
        frametitle=Case of invalid $\Pi$--N dual measures,
        frametitleaboveskip=4pt,
        skipabove=4pt,
        ]\small
\noindent
\begin{description}[leftmargin=!, labelwidth=\widthof{\bfseries Case 2:}]
    \item[Case 1:] $\Pi(A) < 1$ and $\text{N}(A) > 0$. An event cannot be necessary if it is not fully possible.
\end{description}
\end{rhoenv}

\subsubsection*{Summary of Valid $\Pi$--$\textnormal{N}$ Space}

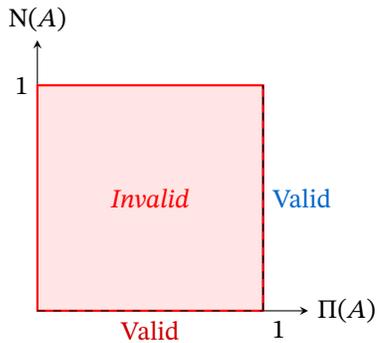
\begin{figure}[h!]
    \centering
    \begin{tikzpicture}[scale=3, >=stealth]
        \draw[->] (0,0) -- (1.2,0) node[right] {$\Pi(A)$};
        \draw[->] (0,0) -- (0,1.2) node[above] {$\text{N}(A)$};
        
        \draw[thick, color=possibilityColor] (1,0) -- (1,1) node[midway, right, color=possibilityColor] {Valid};
        \draw[thick, color=necessityColor] (0,0) -- (1,0) node[midway, below, color=necessityColor] {Valid};
        
        \fill[red!20, opacity=0.5] (1,0) -- (1,1) -- (0,1) -- (0,0) -- cycle;
        \draw[thick, red] (0,0) rectangle (1,1);
        \node at (0.5,0.5) [red] {\textit{Invalid}};
        
        \node[below right] at (1,0) {$1$};
        \node[left] at (0,1) {$1$};
        
        \draw[dashed] (1,0) -- (1,1);
        \draw[dashed] (0,0) -- (1,0);
    \end{tikzpicture}
    \caption{Schematic of valid regions in $\Pi$--N space. Adapted from \citet{Le_Carrer2021-by}.}
    \label{fig:pi_n_space}
\end{figure}

The dual measures can always be represented by a bit and a decimal, also visualised in Fig.~\ref{fig:pi_n_space}.

\begin{table}[h!]
    \centering
    \caption{Summary of Valid Regions in $\Pi$–$\textnormal{N}$ Space}
    \label{tab:summary_valid_space}
    \begin{tabular}{@{}cc@{}}
        \toprule
        \textbf{Condition} & \textbf{Description} \\ 
        \midrule
        $\Pi = 1$ & $\text{N} \in [0,1]$ \\
        $\text{N} = 0$ & $\Pi \in [0,1)$ \\
        \bottomrule
    \end{tabular}\label{tab:valid_regions}
    \vspace{0.5em}
\end{table}

\noindent The valid area in the $\Pi$–$\text{N}$ space is restricted to a reverse-L shape at the perimeter shown in Fig.~\ref{fig:pi_n_space}, mathematically:
\begin{itemize}[noitemsep, topsep=0pt]
    \item A vertical line at $\Pi = 1$ with $\text{N} \in [0,1]$.
    \item A horizontal line at $\text{N} = 0$ with $\Pi \in [0,1)$.
\end{itemize}

In scenarios where multiple outcomes have high possibility but varying necessity measures, this paradigm provides decision-makers a better concept of the range of uncertainties.

\subsubsection*{Representations of uncertainty}
There are multiple ways to derived estimates of uncertainty for the possibility distribution $\pi$:
\begin{itemize}
    \item An measure resembling \textit{information entropy} or a spread of possibility
    \item The width of percentiles when defuzzifying from the area under the curve of aggregated activation of the FIS ruleset (best-case, worst-case scenarios, etc.),
    \item The activation curve ``height" or possibility $\Pi \in [0,1]$, 
    \item After normalisation, necessity $\textnormal{N}$,
    \item The \textit{unsure} value $(1-\max\pi)$ of plausibility missing in the ruleset, given \textit{something} must happen in exhaustive universe $\Omega$.
\end{itemize}

The value of \textit{unsure} indicates explicit weakness in the model for a ruleset that does not cover all inputs. If inputs are clipped to a universe (i.e., range of values on x-axis) then the values bounding $\Omega$ should be adjusted to accommodate all valid input values. Hence, it is implicit an operational FIS assumes the system is stationary. This assumption faces its own risks in predicting air quality, given a constantly changing inventory (i.e., catalogue of industrial sites and their emission rates). The reader may find a longer discussion of non-stationarity in Uinta Basin air quality in \citet{Lyman2022-ac}.

\subsection{Computational Scaling with Grid Spacing in NWP Models}\label{appx:nwp_compute}
The computational cost of a numerical weather prediction (NWP) model increases significantly as the grid spacing ($\Delta x$) is reduced linearly. This increase in demand is driven by several factors: the number of grid points in the horizontal dimensions, the number of vertical levels, and the required reduction in the time step ($\Delta t$) due to the Courant–Friedrichs–Lewy criterion (i.e., information must not move through two grid cells in one timestep). If we assume that the ensemble membership remains constant, the total computational cost $C$ scales approximately as:

\begin{equation}
C \propto \frac{L^2}{\Delta x^3}    
\end{equation}

Where:
\begin{itemize}
    \item $L$ is the domain size (assumed constant),
    \item $\Delta x$ is the horizontal grid spacing,
    \item The exponent of 3 comes from the two spatial dimensions plus the need to increase time-step by the same reduction ratio in $\Delta x$.
\end{itemize}

Thus, halving the grid spacing (e.g., from 2\,km to 1\,km) leads to an approximate 8-fold increase in computational demand \textit{O}(10), while reducing the grid spacing by a factor of 3 (e.g., from 3\,km to 1\,km) results in a $\approx30$-fold increase in computational cost. This represents a significant increase but yet lower than if ensemble size and bottlenecks also scale.

\subsubsection*{Optional Ensemble Scaling}
Finer sampling of future atmospheric states are capturing smaller and smaller circulations. The small width of model truncation means that ensemble membership should increase to capture intricacy in the probability distribution previously not sampled \citep{Tennekes1978-ju}. If ensemble size is also increased to match the finer resolution (to sample more of the model's phase space), we reintroduce the scaling factor related to ensemble members. In this case, the computational cost $C$ scales linearly with the increase in ensemble members.

\subsubsection*{Bottlenecks and Practical Considerations on Supercomputers}

The above scaling relationships assume perfect scaling, but several practical bottlenecks in running NWP models on supercomputers can limit this increase in performance:

\begin{itemize}
    \item \textbf{Tiling and Load Balancing:} Numerical simulations are often divided into tiles that are processed in parallel. In regions of quiescent weather, some tiles may require fewer computations, leading to imbalances where certain tiles finish sooner than others. This introduces latency as the model waits for all tiles to complete before continuing to the next step.
    \item \textbf{I/O Bottlenecks:} Increased resolution means more data needs to be written to storage at each time step, potentially causing bottlenecks in file input/output (I/O) operations.
    \item \textbf{Memory Bandwidth:} The finer resolution results in more memory usage for storing variables at each grid point. Hence, memory access times may become a limiting factor if the memory bandwidth is insufficient, depending on IT architecture.
    \item \textbf{Network Communication Overhead:} As simulations become finer in resolution and require more compute nodes, network latency from information transfer between processors become increasingly significant, particularly when integrating results across multiple tiles.
    \item \textbf{Scalability Issues:} Some algorithms may not scale efficiently beyond a certain number of processors, leading to diminishing returns when adding more computational resources.
\end{itemize}

This neglects further the added resources to run NWP forecasts further into the future, and fast enough (i.e., enough compute power) to be useful---the forecast must arrive quickly for actions to be taken! These factors can cause the actual computational cost to deviate from the ideal scaling relationship, making supercomputing efficiency an important aspect of designing high-resolution NWP models.

\subsection{Seeking a possibilistic measure of gained information}\label{appx:info_gain}

Shannon based information theory on probability theory \citep{Shannon1948-nc}. Imagining a parallel measure to information gain in a possibilistic sense, the measure would similarly increase with improved forecast possibility distributions, measured as the reduction in surprise experienced by the user upon observing the event $A$ after receiving a forecast $f$ of its occurrence \citep{Benedetti2010-sa,Peirolo2011-sl}:

\begin{equation}
    \widetilde{\text{IG}} = \widetilde{I}_{b} - \widetilde{I}_{f}
\end{equation}

where a notational possibility information gain $\widetilde{\text{IG}}$ meesures the increase in a sort of information with a forecast in hand, having previously assumed a distribution $\pi_{b}$. Here, $b$ and $f$ represent baseline and forecast quantities, with a "possibilistic information" content denoted by $\widetilde{I}$. This is a notional measure that parallels information entropy, specifically self-entropy or information content of an object---and as before, the surprise a user experiences observing $A$:

\begin{equation}
    H = -\log_2 f
\end{equation}

This is identical to the Ignorance scoring rule \citep{Roulston2002-eq}. The parallel of entropy measures the uncertainty across the possibility distribution, for instance, in the same manner as for probability forecasts. A confident forecast is peaked, its height dictated by the shape that meets the strictly defined area under the curve. In the possibility view, there is no such restriction. A distribution with a maximum less than 1 (subnormal) as an output of activations allows conflicting or missing information as a feature, not a bug, of the FIS.

It remains an open question how to form a possibility scoring rule that evaluates the acquisition or loss of information by acting on a received forecast $\pi$.


\bibliographystyle{ametsocV6}  
\bibliography{manuscript}  


\end{document}